\documentclass[twocolumn,showpacs,preprintnumbers,amsmath,amssymb,
 aps,
 prb,
 lengthcheck,%
]{revtex4-1}
\usepackage{graphicx}
\usepackage[english]{babel} 
\usepackage[utf8]{inputenc}
\usepackage{hyperref}

\makeindex
\def\>{\right\rangle}
\def\<{\left\langle}
\def\be{\begin{equation}}
\def\ee{\end{equation}}
\def\ba{\begin{array}{l}}
\def\ea{\end{array}}
\def\beq{\begin{eqnarray}}
\def\eeq{\end{eqnarray}}

\usepackage{xcolor}

\begin{document}

 
\title{Theory of non-equilibrium  noise in general multi-terminal superconducting\\ hydrid devices: 
application to multiple Cooper pair resonances} 

\author{R. Jacquet,$^1$ A. Popoff,$^{1,3}$ K.-I. Imura,$^2$ J. Rech,$^1$ T. Jonckheere,$^1$  L. Raymond,$^1$ A. Zazunov,$^4$ 
 and T. Martin$^1$}
\affiliation{$^1$ Aix Marseille Univ, Université de Toulon, CNRS, CPT, Marseille, France}
\affiliation{$^2$  Department of Quantum Matter, AdSM, Hiroshima University, 739-8530, Japan}
\affiliation{$^3$ Collège Tinomana Ebb de Teva I Uta, BP 15001 - 98726 Mataiea, Tahiti, French Polynesia}
\affiliation{$^4$ Institut f\"ur Theoretische Physik, Heinrich Heine Universit\"at, D-40225 D\"usseldorf, Germany}
\date{\today}

\begin{abstract}
We consider the out-of-equilibrium behavior of a general class of mesoscopic devices composed of several superconducting or/and normal metal leads separated by quantum dots. Starting from a microscopic Hamiltonian description, we provide a non-perturbative approach to quantum electronic transport in the tunneling amplitudes between dots and leads: using the equivalent of a path integral formulation, the lead degrees of freedom are integrated out in order to compute both the current and the current correlations (noise) in this class of systems, in terms of the dressed Green's function matrix of the quantum dots. In order to illustrate the efficiency of this formalism, we apply our results to the ``all superconducting Cooper pair beam splitter'', a device composed of three superconducting leads connected via two quantum dots, where crossed Andreev reflection operates Cooper pair splitting. Commensurate voltage differences between the three leads allow to obtain expressions for the current and noise as a function of the Keldysh Nambu Floquet dressed Green's function of the dot system. This voltage configuration allows the occurrence of non-local processes involving multiple Cooper pairs which ultimately lead to the presence of non-zero DC currents in an out-of-equilibrium situation. We investigate in details the results for the noise obtained numerically in the specific case of opposite voltages, where the transport properties are dominated by the
so called ``quartet processes'', involving the coherent exchange of two Cooper pairs among all three superconducting terminals. We show that these processes are noiseless in the non-resonant case, and that this property is also observed for other voltage configurations. When the dots are in a resonant regime, the noise characteristics change qualitatively, with the appearance of giant Fano factors.
\end{abstract} 


\maketitle

\section{Introduction} 

In a mesoscopic device involving superconducting contacts/reservoirs, transport properties are largely influenced by Andreev reflection in the subgap voltage regime.~\cite{andreev64} This fundamental process amounts to an electron being reflected into a hole (or vice-versa), the difference of charge being absorbed by the creation (or destruction) of a Cooper pair (CP) inside the BCS ground state of the superconductor. In a Josephson junction between two superconductors,~\cite{tinkham2004} this microscopic process explains (i) the DC Josephson supercurrent, an equilibrium dissipationless phenomenon which depends on the phase difference between the two superconducting reservoirs, (ii) the AC Josephson effect which depends on the voltage applied across the junction and the Shapiro DC steps obtained when adding an RF irradiation, (iii) the pair-assisted quasiparticle transport for subgap voltages, a phase insensitive non-equilibrium dissipative phenomenon involving multiple Andreev reflections (MAR).~\cite{prb27_btko,prb54_cuevas,prl82_cuevas,prl74_bratus} 

In more involved multi-terminal superconducting hybrid structures, a non-local version of this process may arise. This has been intensively studied in the context of the Cooper pair beam splitter (CPBS),\cite{epjb24_lesovik,prb63_recher,prb70_sauret}  a three lead device composed of a single superconducting lead connected via two quantum dots to two normal metal leads. There, in addition to (direct) Andreev reflection, the two constituent electrons of a Cooper pair can be transferred to the two normal metal leads as a non-local entangled pair\cite{prb66_chtchelkatchev,prb72_sauret} via a process called crossed Andreev reflection (CAR).

\begin{figure}
\centering
\includegraphics[width=0.4\textwidth]{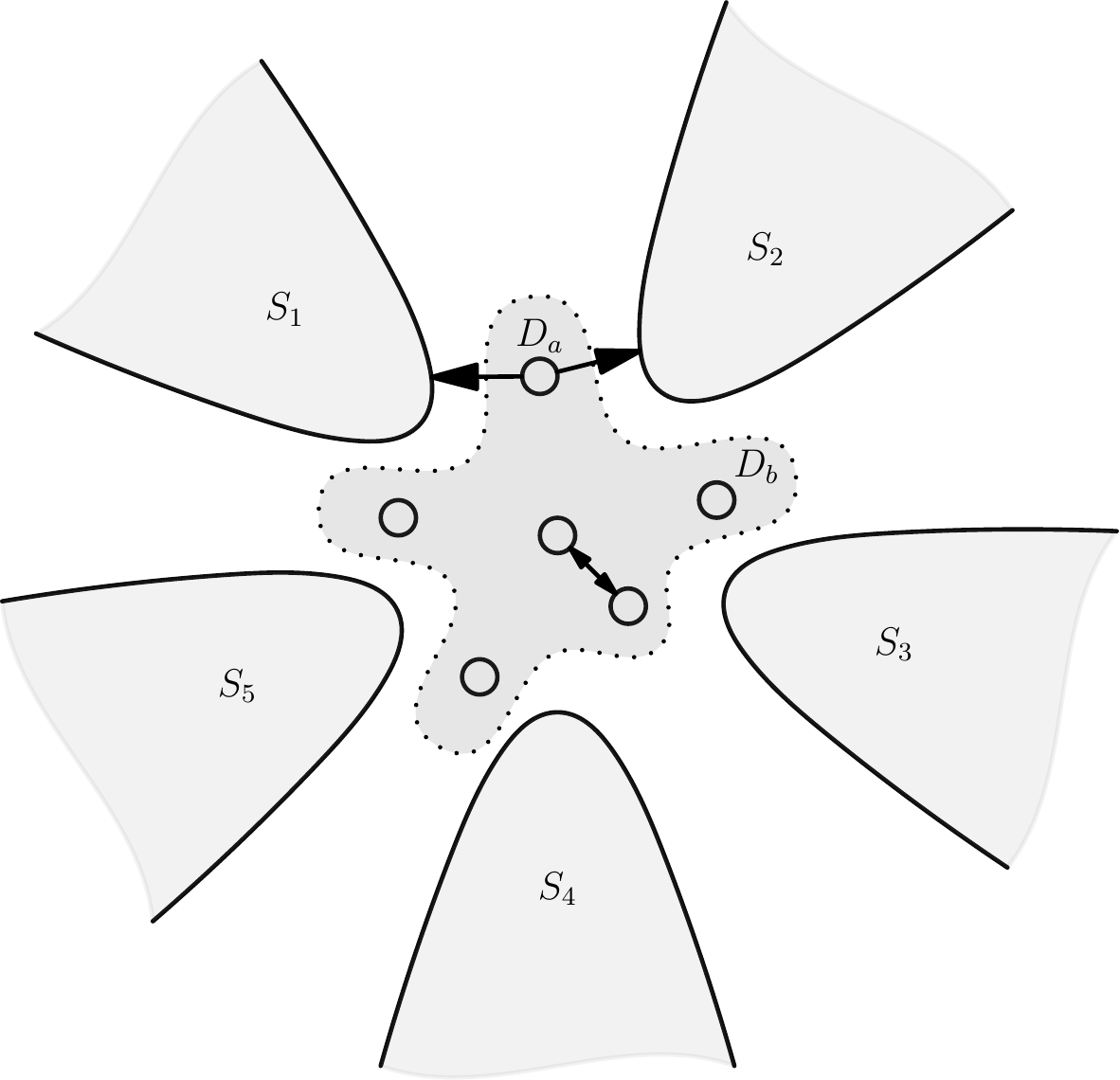}
\caption{Sketch of a general multi-terminal mesoscopic device (with either superconducting or normal metal leads, or both) described by the present formalism: superconducting or normal metal leads are connected to an ``array'' of quantum dots. Electrons can either hop from a quantum dot to one of the leads, or they can alternatively tunnel to another dot. Although no direct tunneling amplitude between the leads is present in this theoretical model, we argue in the text that such processes can  effectively be included by tuning appropriately both the quantum dot level and its hopping amplitude to the leads.}\label{fig_general}
\end{figure}

The first purpose of this paper is to present a general theoretical framework for computing the current and noise characteristics of a general class of mesoscopic devices composed of an arbitrary number of  biased superconducting or normal metal leads,  which are separated by quantum dots. Upon integrating the quadratic degrees of freedom of the leads and applying Wick's theorem, analytical expressions for the noise are obtained in terms of the dressed dot Green's function matrix in the Keldysh Nambu space. To our knowledge, such a general theoretical framework has so far only been considered for the computation of current, in a specific situation: the ``all superconducting Cooper pair beam splitter'' (ASCPBS), studied in Ref. \onlinecite{prb87_jonckheere}, which is the equivalent of the CPBS albeit with all three leads being superconductors.  As is now well established, a noise diagnosis allows to gain further information compared to the current characteristics: particularly relevant is the monitoring of Fano factors (ratios between the noise and the current). We stress that this formulation of quantum transport in superconducting/normal metal hybrid devices can be readily used to study a vast class of systems, as it only requires the (numerical) solution of the Dyson equation of the dressed single particle Green's function of the quantum dot system - the quantum dot ``array''.

In the second part of this paper, as an application of our formalism, we consider the computation of the noise characteristics of the ASCPBS , where in the subgap regime, Andreev reflection, crossed Andreev reflection, and multiple Andreev reflections constitute the basic transport processes. In addition, the voltage differences between the three leads of this ASCPBS are tuned to a commensurate voltage configuration dubbed ``Multiple Cooper pair resonance" (MCPR),\cite{prb87_jonckheere} allowing a dissipationless Josephson-like signal for the current in the subgap regime, despite the fact that the system is driven far from equilibrium. The application of the general formalism developed here is thus to monitor the behavior of the noise as the subgap voltage differences are decreased. In some specific range of parameters where the quantum dot behaves like a quantum point contact, we find that  the low-temperature noise vanishes. Conversely, in the resonant dot regime where the dots operate as energy filters, we find that the subgap voltage noise characteristics gives rise to anomalous giant Fano factors.  

The paper is organized as follows.
Sec.~\ref{sec_model} describes the general class of models which we intend to describe, and the equivalent of a path integral formulation~\cite{prb83_chevallier} for quantum transport is formulated, allowing us to compute both the Josephson current and the current-current correlations in a non-perturbative way in a generalized Fisher-Lee formula\cite{prb23_fisher} for both the current and noise.
In Sec.~\ref{sec_mcpr} we apply our formalism to MCPR in the ASCPBS. After presenting a short historical perspective of selected works dealing with multi-terminal superconducting  hybrid devices, we derive the expressions for the current and noise using our Nambu Keldysh Floquet formalism, and present numerical results in two experimentally relevant regimes. The discussion and conclusions are presented in Secs. ~\ref{sec_discussion} and ~\ref{sec_conclusion}. 

We consider units in which Planck and Boltzmann constants, together with the elementary charge are unity, \textit{i.e.} $\hbar=1$, $k_B=1$, $e=1$.

\section{Microscopic formulation}\label{sec_model}

We consider  a class of mesoscopic systems composed of an arbitrary number of superconducting or normal metal  leads, which are coupled to an array of quantum dots. All superconductors, labeled $j=1,2,...$ are  described by BCS theory with a gap energy $\Delta_j$ (when a given lead is chosen to be a normal metal, this gap is chosen to be zero). Quantum dots are identified with the label $\alpha=a,b,...$ and their energy level is specified by the energy $\varepsilon_\alpha$. There exists no restriction about which quantum dot is coupled to which lead or dot.  The tunneling amplitude between lead $j$ and QD $\alpha$ is denoted as $t_{j\alpha}$, while the direct tunneling between dots  is described by the amplitude $t_{\alpha\beta}=t_{\beta\alpha}$. At first sight, it may seem that a direct coupling (not mediated by quantum dots) is ruled out in this context. This is by no means the case, as we have shown in previous work\cite{prb80_jonckheere} that a system of two superconductors separated by a quantum dot can also describe an adjustable superconducting quantum point contact with arbitrary transmission, provided that the dot level is chosen outside the gaps of the superconductors, and that the dot-superconductor amplitudes are properly tuned.  An example of such devices is depicted in Fig.~\ref{fig_general}. 

\subsection{Total Hamiltonian}\label{subsec_hamiltonian}

The total Hamiltonian of the system can be decomposed into a contribution from the superconducting leads, the one from the QDs (including interdot tunneling) and a final one describing the tunneling between the QD and the leads, namely 
\begin{equation}
H=\sum_{j}H_j+H_D+H_T(t)  .
\end{equation}

The superconducting leads, identified by the label $j$, are represented by the standard BCS Hamiltonian
\begin{equation}
H_j=\sum_k\Psi^\dag_{jk}\left[\left(\frac{k^2}{2m}-\mu\right)\sigma_z+\Delta_j\sigma_x\right]\Psi_{jk}  ,
\end{equation}
where the Pauli matrices $\sigma_x$ and $\sigma_z$ act in Nambu space and we introduced Nambu spinors
\begin{equation}
\Psi_{jk}=
\begin{pmatrix}
\psi_{jk,\uparrow}\\
\psi^\dag_{j(-k),\downarrow}
\end{pmatrix}
  ,
\label{eq_ch3_model_nambu_sc_spinor}
\end{equation}
with $\psi_{jk\sigma}^\dag$ the creation operator for an electron with momentum $k$ and spin $\sigma=\uparrow,\downarrow$ in lead $j$.

Introducing similarly the creation operator for an electron with spin $\sigma=\uparrow,\downarrow$ on dot $\alpha$ as $d_{\alpha\sigma}^\dag$, the Hamiltonian $H_{D_\alpha}$ of quantum dot $\alpha$ reads 
\begin{equation}
H_{D_\alpha}=\epsilon_\alpha\sum_\sigma d^\dag_{\alpha\sigma}d_{\alpha\sigma} ,
\end{equation}
and the tunneling Hamiltonian between the QDs $a$ and $b$ is given by
\begin{equation}
H_{D_aD_b}=t_{ab}\sum_\sigma d^\dag_{a\sigma}d_{b\sigma}+\text{H.c.} 
\end{equation}
Introducing Nambu spinors of the form
\be
d_\alpha=\left(
\begin{matrix}
d_{\alpha\uparrow}\\
d_{\alpha\downarrow}^\dag
\end{matrix}
\right) ,
\ee
and collecting them for each dot $\alpha$ into a $2N_D$-component Nambu-dot spinor ($N_D$ is the number of quantum dots in the system) as
\be
\tilde{d}=\left(
\begin{matrix}
d_a\\
d_b\\
d_c\\
\vdots 
\end{matrix}
\right)  ,
\label{eq_ch3_model_nambu_dot_spinor}
\ee
the Hamiltonian of the double QD can be conveniently rewritten as
\begin{equation}
H_D=\sum_\alpha H_{D_\alpha}+ \sum_{\alpha\beta}H_{D_\alpha D_\beta}=\tilde{d}^\dag\,h_D\sigma_z\,\tilde{d}  ,
\end{equation}
where the Pauli matrix $\sigma_z$ acts in Nambu space and the matrix
\be
h_D=
\begin{pmatrix}
\epsilon_a&t_{ab}&t_{ac}& \dots \\
t_{ba}&\epsilon_b&t_{bc}& \dots \\
t_{ca}&t_{cb}&\epsilon_c & \dots \\
\vdots & \vdots & \vdots & \ddots
\end{pmatrix}  ,
\ee
is defined in dot space. 

Finally, the tunneling Hamiltonian between leads and QDs is written in terms of Nambu spinors according to
\begin{equation}
H_T(t)=\sum_{jk\alpha}\Psi^\dag_{jk}\,{\cal T}_{j\alpha}(t)\,d_\alpha+\text{H.c.}
\end{equation}
where applying a Peierls substitution, we gauged the external bias away from the lead Hamiltonian and into the tunneling constants, therefore introducing ${\cal T}_{j\alpha}(t)=t_{j\alpha}\,\sigma_z\,\text{e}^{i\sigma_z\int_{-\infty}^t V_j \text{d}t}$. 

\subsection{Green's functions in the Keldysh formalism}\label{ch3subsec2_green}

In order to calculate thermodynamic averages of operators in an out-of-equilibrium system, the Keldysh time contour $C$ is introduced:~\cite{rammer_smith} it goes from $-\infty$ to $+\infty$ ($+$ forward branch) and goes back to $-\infty$ ($-$ backward branch). The time ordering operator along this contour is denoted as $T_C$. We introduce the $4 N_D$-component Nambu-dot-Keldysh spinors collecting Nambu-dot spinors~\eqref{eq_ch3_model_nambu_dot_spinor} evaluated on the two different branches of the Keldysh time contour 
\be
\check{d}=\
\begin{pmatrix}
\tilde{d}^{+}\\
\tilde{d}^{-}
\end{pmatrix}  .
\ee
The bare Green's functions of the QDs (in the absence of tunneling between dots and superconducting leads) reads  
\begin{equation}
\check{G}_0(t,t')=-i\left\langle T_C\left\{\check{d}(t)\check{d}^\dag(t')\right\}\right\rangle_0  .
\end{equation}
The quantum mechanical averaging is performed with respect to the Hamiltonian without tunneling
\be
\left\langle\dots\right\rangle_0=\frac{\textrm{Tr}\left\{e^{-\beta H_0}\dots\right\}}{\textrm{Tr}\left\{e^{-\beta H_0}\right\}}\quad\text{where}~ H_0=\sum_j H_j+H_D  .
\ee
QD and superconducting degrees of freedom are coupled with the time-dependent tunneling Hamiltonian $H_T(t)$ and the Green's function dressed by this tunneling reads
\begin{equation}
\check{G}(t,t')=-i\left\langle T_C\left\{S(\infty)\,\check{d}(t)\check{d}^\dag(t')\right\}\right\rangle_0  ,
\end{equation}  
where $S(\infty)$ is the evolution operator along the Keldysh contour
\begin{equation}
S(\infty)=T_C\exp\left\{-i\int_C\text{d}t\,H_T(t)\right\}  .
\end{equation}
Note that for the general class of systems considered here, this Greens function is a $4N_D\times  4N_D$ matrix in Nambu-dot-Keldysh space. 

\subsection{Self energy of the quantum dots}\label{ch3subsec2_self}

The evolution operator when averaged over the lead degrees of freedom takes the form
\begin{align}
\left\langle S(\infty)\right\rangle_\text{leads} = T_C\exp & \left[-i \int_{- \infty}^{+ \infty}
\int_{- \infty}^{+ \infty}\text{d}t_1\,\text{d}t_2 \right.  \nonumber \\
& 
\qquad \check{d}^\dag(t_1)\check{\Sigma}_T(t_1,t_2)\check{d}(t_2) \Bigg]  ,
\end{align}
involving a total self-energy $\check\Sigma_T=\sum_j\check{\Sigma}_j$ which also takes the form of a matrix in Nambu-dot-Keldysh space. Each lead self-energy $\check\Sigma_j$ can be viewed as a set of Nambu-Keldysh matrices given by
\begin{equation}
\left[\check{\Sigma}_{j}\right]_{\alpha\beta}(t_1,t_2)={\cal T}_{j\alpha}^\dag(t_1)\tau_{z}\hat{g}_{j}(t_{1}-t_{2})\tau_{z}{\cal T}_{j\beta}(t_{2})  ,
\label{eq:Sigmaj}
\end{equation}
and corresponding to the $N_D^2$ possible matrix elements in dot space. There, the new set of Pauli matrices $\tau_{x,y,z}$ acts in Keldysh space, and we introduced
\begin{equation}
\hat{g}_{j}(t-t^\prime)=-i\sum_{k}\left\langle T_C\left\{\hat{\Psi}_{jk}(t)\hat{\Psi}_{jk}^\dag(t')\right\}\right\rangle_0 ,
\end{equation}
as the bare local Green's function of the superconducting lead $j$ at the site of tunneling. It involves the Nambu-Keldysh spinors which collect the Nambu spinors~\eqref{eq_ch3_model_nambu_sc_spinor} evaluated on the two different branches of the Keldysh time contour as
\be
\hat{\Psi}_{jk}=
\begin{pmatrix}
\Psi_{jk}^+\\
\Psi_{jk}^-
\end{pmatrix}  .
\ee

In order to carry out some of the upcoming calculations, it is useful to perform a rotation in Keldysh space going from the $+/-$ basis to the so-called $RAK$ basis according to
\begin{equation}
\begin{pmatrix}
g_j^R&g_j^K\\
0&g_j^A
\end{pmatrix}=L\tau_z\hat{g}_jL^{-1}\quad\text{with}~
L=\frac{1}{\sqrt{2}}
\begin{pmatrix}
1 & -1 \\
1 & 1
\end{pmatrix} .
\end{equation}
This defines a new set of Green's functions corresponding to the retarded ($R$), advanced ($A$) and Keldysh ($K$) components. In the present case of BCS superconductors, these are given by
\begin{equation}
\left\{
\begin{aligned}
&g_j^{R,A}(\omega)=\pi\nu(0)\,\frac{\omega+\Delta_j\sigma_{x}}{i\zeta_j^{R,A}(\omega)}  ,\\
&g_j^K(\omega)=\left[1-2f(\omega)\right]\left[g_j^R(\omega)-g_j^A(\omega)\right]  ,
\end{aligned}
\right.
\label{eq:gRAK}
\end{equation}
where $\nu(0)$ is the density of states of the lead in the normal metal regime at the Fermi level, $f(\omega)$ is the Fermi distribution and we introduced the functions
\begin{align}
\zeta_j^{R,A}(\omega)=
&\pm\text{sign}(\omega)\sqrt{\omega^2-\Delta_j^2}\,\Theta\left(|\omega|-\Delta_j\right)\notag\\
&+i\sqrt{\Delta_j^2-\omega^2}\,\Theta\left(\Delta_j-|\omega|\right)  .
\label{eq:zetaRA}
\end{align}

\subsection{Current statistics}\label{sec_corr}

\subsubsection{Current operator and average}\label{ch3subsec3_josephson}

The current operator from QD $\alpha$ into the lead $j$ reads 
\begin{equation}
I_{j\alpha}(t)=i\sum_k\Psi^\dag_{jk}\,\sigma_z{\cal T}_{j\alpha}(t)\,d_\alpha+\text{H.c.}
\end{equation}
As the average current does not depend on the branch of the Keldysh contour, it is convenient to introduce counting fields $\eta_{j\alpha}(t)$ in the tunneling amplitudes according to
\be
{\cal T}_{j\alpha}(t)\rightarrow{\cal T}_{j\alpha}(t)\,\text{e}^{i\sigma_z\tau_z\eta_{j\alpha}(t)/2}  ,
\ee 
so that the evolution operator becomes $S(\infty)\rightarrow S(\infty,\eta)$. The average current can then be computed through the functional differentiation
\begin{equation}
\left\langle I_{j\alpha}\right\rangle(t)=i\,\frac{1}{Z[0]}\left.\frac{\delta Z\left[\eta\right]}{\delta\eta_{j\alpha}(t)}\right|_{\eta=0} ,
\end{equation}
where $Z[\eta]=\left\langle S(\infty,\eta)\right\rangle_0~$. Performing the differentiation explicitly, we obtain a Meir-Wingreen type formula~\cite{prl68_meir} for the average current as the following $\alpha\alpha$ diagonal element in dot space
\begin{align}
\left\langle I_{j\alpha}\right\rangle(t)=\frac{1}{2}\,\text{Tr}^{(NK)}\Bigg\{\sigma_z\tau_z\int^{+\infty}_{-\infty}\text{d}t'&\Big[\check{G}(t,t')\check{\Sigma}_j(t',t)\notag\\
-&\check{\Sigma}_j(t,t')\check{G}(t',t)\Big]^{\alpha\alpha}\Bigg\}  ,
\end{align}
where $\text{Tr}^{(NK)}$ denotes the trace in Nambu-Keldysh space. 

\subsubsection{Current correlations}\label{ch3subsec3_noise}

In full generality, we need to compute the unsymmetrized current-current correlator defined as
\begin{align}
S_{i\alpha,j\beta}(t,t')  = \left\langle I_{i\alpha}(t)\,I_{j\beta}(t')\right\rangle  -  \left\langle I_{i\alpha}(t)\right\rangle \left\langle I_{j\beta}(t')\right\rangle  .
\label{eq:currcorr}
\end{align}
A convenient way of doing so consists in introducing new counting fields~\cite{prb83_chevallier}  $\eta_{j\alpha s}(t)$ where $s=\pm$ now specifies the branch of the Keldysh contour. 
The tunneling amplitudes are then redefined following this prescription as
\begin{equation}
{\cal T}_{j\alpha}(t)\rightarrow{\cal T}_{j\alpha}(t)\,\text{e}^{i\sigma_z\sum\limits_s\pi_s\eta_{j\alpha s}(t)}  ,
\end{equation}
where we defined the following matrices in Keldysh space to project onto a given branch of the contour
\begin{equation}
\pi_\pm = \frac{\tau_z \pm 1}{2}  .
\end{equation}

The current correlations are then computed through second order functional differentiation as
\begin{equation}
\left\langle I^-_{i\alpha}(t)\,I^+_{j\beta}(t')\right\rangle=-\frac{1}{Z[0]}\left.\frac{\delta^2 Z[\eta]}{\delta\eta_{i\alpha-}(t)\,\delta\eta_{j\beta+}(t')}\right|_{\eta=0}  ,
\end{equation}
where $Z[\eta]=\left\langle S(\infty,\eta)\right\rangle_0~$. 
\begin{widetext}
Performing this differentiation, using Wick theorem, and carrying out the partial trace over Keldysh space, it can eventually be expressed in terms of $RAK$ components as~\cite{prb99_bathelier}
\begin{align}
&S_{i\alpha,j\beta}(t,t')=-\frac{1}{2}\,{\rm{}Re}\int^{+\infty}_{-\infty}\text{d}t_1\int^{+\infty}_{-\infty}\text{d}t_2\,\notag\\
&\times\text{Tr}^{(N)}\left\{\sigma_{z}\left(\tilde{\Sigma}_{i}^{K}\tilde{G}^{A}+\tilde{\Sigma}_{i}^{R}\tilde{G}^{K}-\tilde{\Sigma}_{i}^{A}\tilde{G}^{A}+\tilde{\Sigma}_{i}^{R}\tilde{G}^{R}\right)^{\alpha\beta}_{(t,t_{1})\circ(t_{1},t^{\prime})}
\sigma_{z}\left(\tilde{\Sigma}_{j}^{K}\tilde{G}^{A}+\tilde{\Sigma}_{j}^{R}\tilde{G}^{K}+\tilde{\Sigma}_{j}^{A}\tilde{G}^{A}-\tilde{\Sigma}_{j}^{R}\tilde{G}^{R}\right)^{\beta\alpha}_{(t^{\prime},t_{2})\circ(t_{2},t)}\right.\notag\\
&\left.-\sigma_{z}\left(\tilde{\Sigma}_{i}^{R}\tilde{G}^{R}\tilde{\Sigma}_{j}^{K}+\tilde{\Sigma}_{i}^{K}\tilde{G}^{A}\tilde{\Sigma}_{j}^{A}+\tilde{\Sigma}_{i}^{R}\tilde{G}^{K}\tilde{\Sigma}_{j}^{A}
-\tilde{\Sigma}_{i}^{A}\tilde{G}^{A}\tilde{\Sigma}_{j}^{A}+\tilde{\Sigma}_{i}^{R}\tilde{G}^{R}\tilde{\Sigma}_{j}^{R}\right)^{\alpha\beta}_{(t,t_{1})\circ(t_{1},t_{2})\circ(t_{2},t^{\prime})}
\sigma_{z}\left(\tilde{G}^{K}+\tilde{G}^{A}-\tilde{G}^{R}\right)^{\beta\alpha}_{(t^{\prime},t)}\right\}\notag\\
&- \frac{\delta_{ij}}{4}\text{Tr}^{(N)}\left\{ \sigma_z \left(\tilde{\Sigma}^A_i - \tilde{\Sigma}^R_i - \tilde{\Sigma}^K_i\right)^{\alpha \beta}_{(t,t')}  \sigma_z \left(\tilde{G}^A - \tilde{G}^R + \tilde{G}^K \right)^{ \beta \alpha}_{(t',t)}\right.   \notag\\ 
&\hspace{6em} + \left.\sigma_z \left(-\tilde{G}^A + \tilde{G}^R + \tilde{G}^K \right)^{\alpha \beta}_{(t,t')} \sigma_z \left(-\tilde{\Sigma}^A_i + \tilde{\Sigma}^R_i - \tilde{\Sigma}^K_i\right)^{ \beta \alpha}_{(t',t)} \right\}
  ,
\label{real_time_noise_RAK}
\end{align}
where $\circ$ stands for a convolution product in time.
\end{widetext}

Note that as the self energies are proportional to squares of tunneling amplitudes, the  auto-correlation noise always dominates with respect to the crossed correlation noise at low transparencies: the $\delta_{ij}$ term is the dominant one in Eq.~\eqref{real_time_noise_RAK} when lowest order perturbation theory is operated. 

This concludes the first task of this paper. The current and noise have been expressed in terms of the dressed Green's function matrix elements, for an arbitrary mesoscopic system composed of an array of quantum dots coupled in full generality to a set of superconducting or normal metal leads in the spirit of a generalized Fisher-Lee formula\cite{prb23_fisher} applicable to (time dependent) superconducting systems. In order to make further progress, one needs to solve the corresponding Dyson's equation numerically for a specific device. 

If the system contains only a single superconductor, only (single) Andreev reflection and quasiparticle transmission to/from this superconductor  can occur: a stationary current flows in this system, and current correlations depend only on the time difference $t-t'$.  The spectral density of noise at finite frequencies can be directly obtained by a Fourier transform with respect to $t-t'$ . The results of Refs. \onlinecite{prb83_chevallier,prb85_rech}, which consider the noise crossed correlation is the CPBS belong to this category. 

However, the next goal of this work is precisely to consider systems with 3 superconductors, where the currents flowing in the device contain ac and dc components, and where the real time current correlator depends separately on the two times $t$ and $t'$. 

\begin{figure}
\centering
\includegraphics[width=0.4\textwidth]{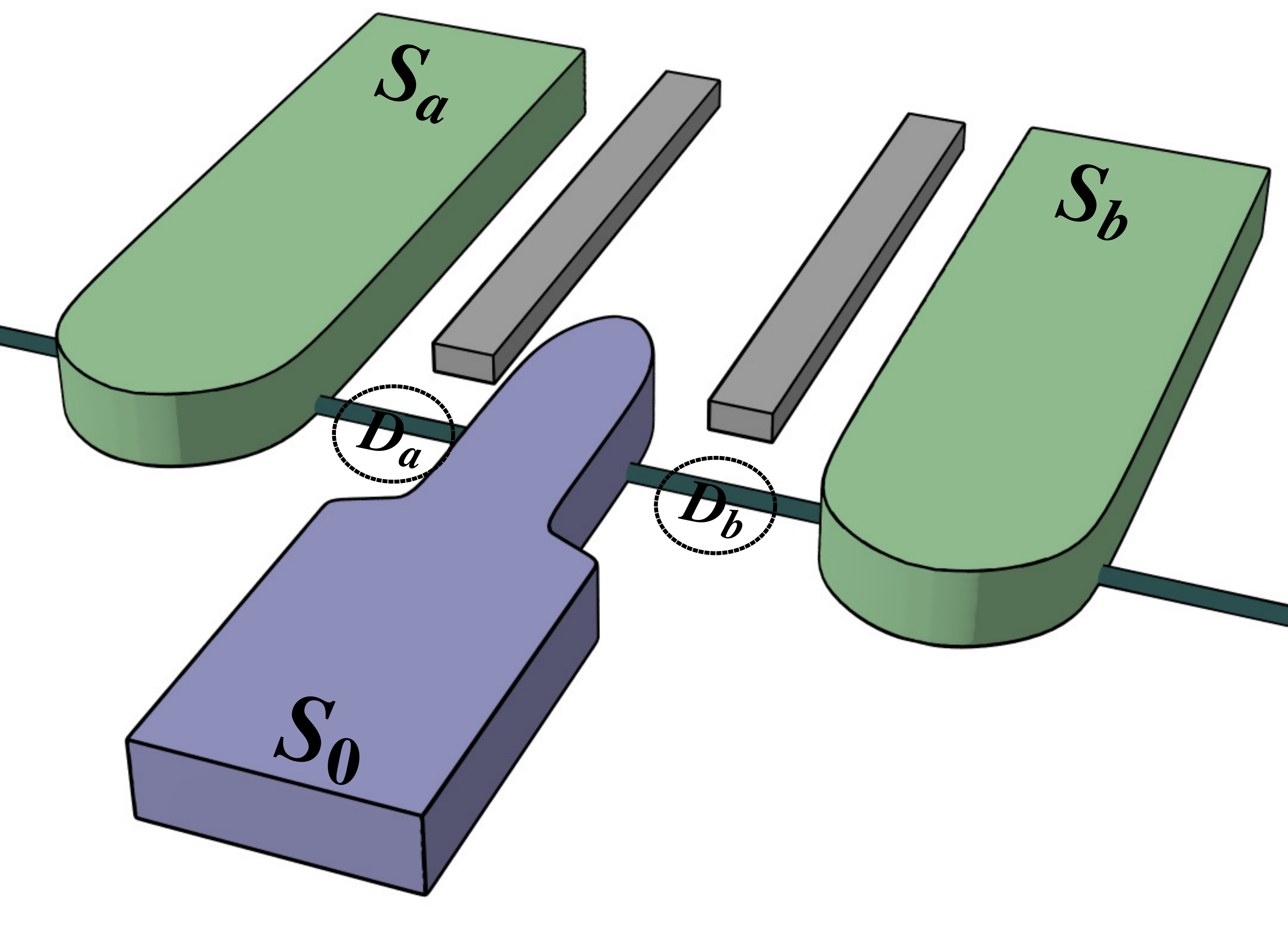}
\caption{Three superconductors designed in an ASCPBS.~\cite{prb87_jonckheere} The central electrode $S_0$ is grounded while the lateral ones $S_a$ and $S_b$ are biased with voltages $V_a$ and $V_b$. Two quantum dot nanowires $D_a$ and $D_b$, with energies $\epsilon_a$ and $\epsilon_b$ which can be tuned by gates (in gray), bridge the central superconductor to the two lateral ones. The distance between the two dots is comparable to the coherence length.}\label{fig_bijunc}
\end{figure}

\section{Application to Multiple Cooper pair resonances}
\label{sec_mcpr}

In this section, we study the setup presented in Fig.~\ref{fig_bijunc} for multiple Cooper pair splitting in a central superconductor, and recombination (as Cooper pairs) in two other (voltage biased) superconducting leads. 
We thus consider a central grounded superconducting electrode $S_0$ coupled to two lateral superconducting leads $S_a$ and $S_b$ via two quantum dots (QDs) $D_a$ and $D_b$ as illustrated in Fig.~\ref{fig_bijunc}. All superconductors labeled $j=0,a,b$ are described by BCS theory with a gap energy $\Delta_j$. Each QD labeled $\alpha=a,b$ characterized by an energy level $\epsilon_\alpha$ is coupled to the central lead and to (only) one lateral superconductor: the tunneling amplitude between lead $j$ and QD $\alpha$ is denoted as $t_{j\alpha}$ and the direct inter-dot coupling is neglected  ($t_{ab}=t_{ba}^*=0$) for simplicity, and to optimize CAR processes.
The two lateral leads are biased with a voltage $V_j$ measured with respect to the chemical potential of the central superconducting electrode with $V_0=0$. The width of $S_0$ is assumed to be smaller than the superconducting coherence length, so that CAR processes can operate, splitting pairs from $S_0$ and distributing electrons on both QDs (as Cooper pairs). 

\subsection{Historical perspective}
\label{sec:history}

Early works~\cite{prl102_duhot,prb82_houzet,prb82_chtch,prl107_kaviraj} attempting to describe the transport properties of three-terminal all-superconducting junctions have focused on the so-called incoherent regime.~\cite{prb62_bezu} 
Further investigations made clear that in the coherent regime, CAR processes~\cite{prl74_byers_flatte,pla220_martin,prb53_anantram_datta,apl76_deutscher_feinberg,jpcm13_melin} would allow the correlated motion of CPs originating from all three superconductors, leading to interesting signatures in subgap transport.~\cite{prb87_jonckheere}  It was indeed realized that the interference of MAR processes taking place at different interfaces of the three-terminal device would lead to the intriguing possibility of phase-sensitive dissipative transport. The most fascinating feature, however, is the appearance of so-called multiple Cooper pair resonances (MCPR) which resemble some form of Andreev bound states delocalized over all superconductors, leading to a dissipation-less phase-dependent Josephson-like current in a non-equilibrium situation. 

The lowest order MCPR results in the entanglement of two CPs, a process referred to as the ``quartet'' process and first envisioned (at least in these terms) in the equilibrium calculations of Ref.~\onlinecite{prl106_freyn}. In another formulation, these resonances were predicted as voltage-induced (fractional) Shapiro steps.~\cite{prb75_cuevas} Anomalies observed recently in the electronic subgap transport of an all-superconducting device~\cite{prb90_pfeffer,IEEE_duvauchelle} could meet an interpretation in terms of quartet resonances. 

Here, we wish to determine the noise characteristics of MCPR. Indeed, the correlations between currents flowing in two different leads (noise crossed correlations) of a multi-terminal setup can be measured in order to probe non-local correlations, such as the ones resulting from CAR processes. In particular, the sign of such current correlations has been used as a way to sharpen our understanding of mesoscopic devices involving superconductors from both the theoretical and experimental point of view.~\cite{pla220_martin,prb53_anantram_datta,epjb12_torres,nat_com_heiblum,prl102_duhot,prb82_freyn,prl107_kaviraj} In a setup where a single lead is biased, low voltage positive noise crossed correlations were predicted and ascribed to MARs.~\cite{physicae_riwar_reprint} In a setup consisting of a single quantum dot connected to three terminals, noise crossed correlations were also investigated using perturbative calculations~\cite{prb93_melin} and quartets were  shown to have a decreasing noise signal at low subgap voltages for a non-resonant dot, as opposed to the resonant case where a phase-sensitive noise was predicted. 
The study of out-of-equilibrium noise in multi-terminal superconducting junctions in the coherent regime is particularly interesting, as there is still a need to quantify the evolution of current correlations between different terminals, as a function of the phase and the voltage biases.
In fact, the measurement of positive noise crossed correlations in three-terminal all-superconducting devices~\cite{pnas_cohen} has been reported recently, in agreement with the differential conductance anomaly ascribed to quartets. 

\subsection{Short description of MCPR physics} \label{sec:description}

Three-terminal devices involving a superconducting (S) source connected to two normal (N) metallic leads have been extensively studied,~\cite{prl74_byers_flatte,pla220_martin,prb53_anantram_datta,apl76_deutscher_feinberg,jpcm13_melin,prb84_burset} mostly because they offer the possibility to generate two-particle entanglement~\cite{epjb24_lesovik,prb63_recher,nanotech14_bouchiat,prl91_samuelsson} by extracting a split Cooper pair from the BCS condensate, which granted them the name ``N-S-N Cooper pair splitters''. Microscopically, Cooper pair splitting is ensured by the process of crossed Andreev reflection (CAR) which allows an ongoing hole in one normal lead to be reflected as an outgoing electron in the other N lead, making use of the (evanescent) quasiparticle states in S, provided that the separation between the metallic reservoirs is smaller than the coherence length of the superconducting material. Such devices were realized experimentally, with convincing evidence of both nonlocal current and noise.~\cite{prl93_beckmann,prl95_russo,prl97_caddenzimansky,naturenano_cleuziou,nature461_hofstetter,prl104_herrmann,prl107_hofstetter,nat_com_heiblum}

\begin{figure}
\centering
\includegraphics[width=0.45\textwidth]{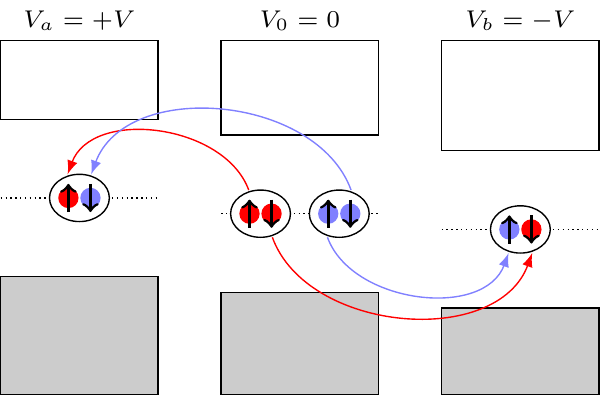}
\caption{Energy diagram of the quartet production mechanism. For opposite applied voltages $V_a = - V_b = V$, a process involving two crossed Andreev reflections splits two Cooper pairs from the central grounded electrode, leading to the formation of a four fermion entangled state, with two pairs emitted in the lateral superconductors, one in $S_a$ and one in $S_b$.}\label{fig:quartets}
\end{figure} 

The concept of Cooper pair splitting can be generalized to an all-superconducting device~\cite{prl106_freyn} where an external voltage bias is applied to the lateral superconducting leads, while the central one is grounded. In such a system, single-particle conduction inside the outgoing biased leads is prohibited for subgap voltages. However, CAR processes can still operate and, for properly selected applied voltages, lead to interesting phenomena. Indeed, focusing first on the case of opposite potential biases for simplicity, one can envision a process by which two Cooper pairs originating from the central lead are split into entangled nonlocal pairs by virtue of a double crossed Andreev reflection, their constituent electrons ultimately recombining as newly formed Cooper pairs in the lateral leads. Such a process, shown schematically in Fig.~\ref{fig:quartets}, therefore leads to the coherent transfer of two Cooper pairs, relying on a combination of both direct and crossed Andreev reflections. It ultimately corresponds to the formation of a correlated four fermion state, an object sometimes referred to as ``nonlocal quartet''. More importantly, this is an energy-conserving process which results in signatures in the DC current, which turn out to bear a nontrivial phase and voltage dependence.

Such a process can be extended to involve any even number of CAR processes, a key requirement for the split pairs originating from the central electrode to get recombined in the lateral leads. Quartets are thus easily generalizable to higher-order multiple Cooper pair resonances (MCPR) implicating the phase-coherent transport of $n+m$ pairs from $S_0$ being transferred as $n$ pairs to $S_a$ and $m$ pairs to $S_b$. This, in turn, leads to the appearance of DC Josephson resonances at commensurate voltages, thus satisfying 
\begin{align}
n V_a + m V_b = 0 .
\label{eq:commensurability}
\end{align}

The existence of the quartet resonance, and more generally of the MCPR, can be inferred from a simple phase argument, as proposed in Ref.~\onlinecite{prb87_jonckheere}, resulting in the commensurability condition, Eq.~\eqref{eq:commensurability}, to be satisfied by the applied voltages. In an equilibrium setup, the current-phase relation is obtained from differentiating the Josephson free energy with respect to the superconducting phases $\varphi_j$ associated with each lead $S_j$. 

When external voltages $V_a$ and $V_b$ are applied to the lateral leads $S_j$ ($j = a,b$), the superconducting phases acquire a time dependence, namely $\varphi_a(t)=\varphi_a + 2 V_a t$, and $\varphi_b(t)=\varphi_b + 2 V_b t$, while we assumed $\varphi_0=0$ for the grounded central electrode, so that $\varphi_0(t)= 0$. It follows that focusing on the currents $I_j$ in the lateral leads $S_j$ ($j = a,b$), one is left with
\begin{equation}
I_j(t)=\sum_{p, q\in\mathbb{Z}}I_{j,pq}\sin\left[p \varphi_a(t) + q \varphi_b(t)\right]  .
\label{pheno}
\end{equation}
In the special case where the applied voltages are commensurate, i.e. there exists a pair of integers $(n,m)$ such that $n V_a + m V_b = 0$, one readily sees that the currents $I_j (t)$ now contain a term which is constant in time, corresponding to the component $\left\{ p=n, q=m \right\}$ along with its higher order harmonics. This results in a pure DC component, signaling the existence of a multiple Cooper pair resonance. It depends on the combination of the bare phases $n \varphi_a + m \varphi_b$, which corresponds to the transfer of $n+m$ Cooper pairs from $S_0$, outgoing as $n$ pairs in $S_a$ and $m$ in $S_b$. 

The quartet resonance discussed in this study constitutes the ``lowest order'' MCPR,  observed for $n=m=1$, which in a truncated perturbative treatment in the tunnel amplitudes, constitutes the dominant MCPR. Here, our application to this quartet process is by no means perturbative, and it allows to access transport regimes at high transparency through all junctions. In Ref. \onlinecite{prb87_jonckheere}, the authors performed a systematic study of the amplitude of the DC Josephson current for such higher order processes. However, when the integers $m$ and $n$ identifying the MCPR are instead chosen to be large, as well as when high transparencies are specified, one is limited by numerical power. This is the reason why, as a first application of our noise diagnosis, we focus on the quartet case.    


\subsection{Expressing the current and noise at a MCPR}

\subsubsection{Double Fourier representation and Dyson equation}\label{ch3subsec2_fourier}
\label{sec:double_fourier}

When arbitrary voltages $V_a$ and $V_b$ are applied to the lateral superconducting leads while keeping the central superconducting electrode grounded, two Josephson frequencies $2|V_a|$ and $2|V_b|$ govern the system. In general, they are independent and the QD Green's function is a function of two times $t_1$ and $t_2$ or alternatively of $\tau=t_1-t_2$ and $t=(t_1+t_2)/2$.  However, when the applied voltages are commensurate, $n V_a + m V_b = 0$ with $n,m$ integers, the average time variable $t$ becomes periodic, with period $T=\pi|m/V_a|=\pi|n/V_b|\equiv 2\pi/ \omega_J$ thus defining an effective Josephson frequency $\omega_J$. Note that this condition corresponds precisely to the one allowing for the appearance of MCPR. It then becomes possible to describe the bare and dressed QD Green's functions, $\check{G}_0$ and $\check{G}$, along with the self energies $\check\Sigma_j$, using a single frequency and two harmonics indexes representing harmonics of the Josephson frequency $\omega_J$, in the spirit of Floquet theory.~\cite{prb80_jonckheere} This specific double Fourier transform representation thus allows us to write all Green's functions as frequency-dependent matrices in harmonics space (on top of their Nambu-dot-Keldysh structure)
\begin{align}
\check{G}_{nm}\left(\omega\right)=\frac{\omega_J}{2\pi} & \int \text{d}t_1  {e}^{i(\omega+n \omega_J)t_1} \nonumber \\
  & \times \int \text{d}t_2  {e}^{-i(\omega+m \omega_J)t_2} \check{G}\left(t_1,t_2\right) .
\label{eq:defGnm}
\end{align}

This additional matrix structure in harmonics space offers an important advantage over other formulations as it allows to write the Dyson equation as a simple matrix inversion (albeit in the rather large Nambu-dot-Keldysh-harmonics space)
\begin{equation}
\check{G}(\omega)^{-1}=\check{G}_0(\omega)^{-1}-\check\Sigma_T(\omega)  ,
\end{equation}
which is translated into $RAK$ components as
\begin{align}
&\tilde{G}^{R/A}(\omega)^{-1}=\tilde{G}_0^{R/A}(\omega)^{-1}-\tilde\Sigma_T^{R/A}(\omega)  ,\label{eq_ch3_model_dyson_inversion}\\
&\tilde{G}^K(\omega)=\tilde{G}_0^K(\omega)+\tilde{G}^R(\omega)\tilde\Sigma_T^K(\omega)\tilde{G}^A(\omega)  .
\end{align}

\begin{widetext}
The bare Green's function is not only diagonal in Keldysh space (expressed in the $RAK$ basis) as $\tilde{G}_0^K=0$, but also in harmonics space as, when expressed in the time domain, it only depends on the time difference $\tau = t_1-t_2$.  These diagonal elements can be written as Nambu-dot matrices taking the form
\be
\left[\tilde{G}_0^{R/A}(\omega)^{-1}\right]_{nm}=\delta_{nm}
\begin{pmatrix}
\omega+n \omega_J - \epsilon_a\sigma_z & -t_{ab}\sigma_z \\
-t_{ba}\sigma_z & \omega+ n \omega_J -\epsilon_b\sigma_z 
\end{pmatrix}  .
\ee
Similarly, the lead self-energies can be obtained in this enlarged space using Eqs.~\eqref{eq:Sigmaj}, \eqref{eq:gRAK}-\eqref{eq:defGnm}, and are given by:~\cite{prb73_zazunov_egger}
\begin{equation}
\left[\check{\Sigma}_j(\omega)\right]_{nm}=\Gamma_j
\begin{pmatrix}
\delta_{n m}\,\hat{X}_j(\omega+n \omega_J -V_j) & \delta_{n-2V_j/ \omega_J ,m}\,\hat{Y}_j(\omega+n \omega_J -V_j) \\
\delta_{n+2V_j/ \omega_J ,m}\,\hat{Y}_j(\omega+n \omega_J +V_j) & \delta_{n m}\,\hat{X}_j(\omega+n \omega_J +V_j) \\
\end{pmatrix}  ,
\label{eq_ch3_model_self}
\end{equation}
\end{widetext}
where $\Gamma_j$ is a matrix in dot space with matrix elements $\Gamma_{j\alpha\beta}=\pi\nu(0)t^*_{j\alpha}t_{j\beta}~$, and where $\hat{X}_j$ and $\hat{Y}_j$ are matrices in Keldysh space, with components expressed in the RAK basis as 
\be
\left\{
\begin{aligned}
&X_j^{R/A}(\omega)=-\frac{\Theta(\Delta_j-|\omega|)\,\omega}{\sqrt{\Delta_j^2-\omega^2}}\mp i\,\frac{\Theta\left(|\omega|-\Delta_j\right)|\omega|}{\sqrt{\omega^2-\Delta_j^2}}  ,\\
&X_j^K(\omega)=-2i\,\frac{\Theta\left(|\omega|-\Delta_j\right)|\omega|}{\sqrt{\omega^2-\Delta_j^2}}\tanh\frac{\beta\omega}{2}  , \\
&Y_j^{R,A,K} (\omega)=-\Delta_j\,\frac{X_j^{R,A,K} (\omega)}{\omega}  .
\end{aligned}
\right.
\ee
In all generality, an exact description of the problem would require an infinite number of harmonics. In practice, we can restrict ourselves to a finite subset by introducing a cutoff energy $E_c$, which needs to be much larger than any relevant energy scale of the problem (typically a few times the largest superconducting gap). The dressed dot Green's function $\tilde{G}^{R,A,K}$ is then obtained numerically from Eq.~\eqref{eq_ch3_model_dyson_inversion}, through the inversion of large matrices. The typical size of these objects is dictated by the applied voltage biases through the Josephson frequency, as one needs $N\sim E_c/ \omega_J $ frequency domains of size $ \omega_J $ to properly cover the range of energy up to the cutoff $E_c$. This makes the handling of low-voltage situations particularly time-consuming, and numerically challenging because of resonances which require to perform integrals with a finer resolution on increasing size intervals.

Note that all analytical results can directly be applied to more general multi-terminal superconducting devices, provided that the MCPR condition is satisfied between potential differences with respect to the ground. This is also true for the results of the section below, which are not sample specific.

\subsubsection{Current  harmonics and noise in frequency space}\label{ch3subsec3_noise_mcpr}

As argued earlier, multiple Cooper pair resonances appear when commensurate voltages are applied to the lateral superconducting leads, a condition which also allows to recover a periodic behavior of the transport properties in terms of a single (effective) Josephson frequency $\omega_J$. When the MCPR condition is satisfied, the current admits a Fourier series expansion of the form
\be
\left\langle I_{j\alpha}\right\rangle(t)=\sum_{p\in\mathbb{Z}}\text{e}^{-ip \omega_J t}\,{\cal I}_{j\alpha}^p  ,
\ee
\begin{widetext}
with Fourier coefficients given by~\cite{prb87_jonckheere}
\begin{align}
{\cal I}_{j\alpha}^p=\frac{1}{2}\,\text{Tr}^{(NK)}\!\Bigg\{\sigma_z\tau_z\!\int_0^{\omega_J}\frac{\text{d}\omega}{2\pi}\sum_n\Big[&\check{G}(\omega)\check\Sigma_j(\omega) -\check\Sigma_j(\omega)\check{G}(\omega)\Big]^{\alpha\alpha}_{n,n-p}\Bigg\}  .
\end{align}

Performing the partial trace over Keldysh space, it can be expressed in terms of $RAK$ components as
\begin{equation}
{\cal I}_{j\alpha}^p=\frac{1}{2}\,\text{Tr}^{(N)}\Bigg\{\sigma_z\int_0^{\omega_J}\frac{\text{d}\omega}{2\pi}\sum_n\Big[\tilde{G}^R(\omega)\tilde{\Sigma}^K_j(\omega)+\tilde{G}^K(\omega)\tilde{\Sigma}^A_j(\omega)
-\tilde{\Sigma}^R_j(\omega)\tilde{G}^K(\omega)-\tilde{\Sigma}^K_j(\omega)\tilde{G}^A(\omega)\Big]^{\alpha\alpha}_{n,n-p}\Bigg\}  .
\label{eqch3_corr_current}
\end{equation}
\end{widetext}

From the definition of the current-current correlations, Eq.~\eqref{eq:currcorr}, one can introduce the finite-frequency noise, calculating the Fourier transform with respect to the time difference as
\begin{align}
S^+_{i\alpha,j\beta} (\Omega,t)   &\equiv \int_{-\infty}^{+\infty}\text{d}\tau\,\text{e}^{i\Omega \tau}S_{i\alpha,j\beta}(t,t+\tau)  ,\\
S^-_{i\alpha,j\beta} (\Omega,t)  &\equiv \int_{-\infty}^{+\infty}\text{d}\tau\,\text{e}^{i\Omega \tau}S_{i\alpha,j\beta}(t+\tau,t)  ,
\end{align}
which correspond to the emission and absorption noise. In inductive coupling schemes for the detection of noise, \cite{jetp65_lesovik,prl99_zazunov} at low temperatures for both the device under study and the detector, the emission noise typically dominates. 

Under the MCPR condition these Fourier-transformed correlators recover a periodicity in their average time variable. Like the currents, they contain all harmonics of the Josephson frequency, which motivates the computation of the following time averages~\cite{prl82_cuevas} corresponding to the emission and absorption noise:
\begin{align}
\label{eq:Splus}
\bar{S}^+_{i\alpha,j\beta}(\Omega) &\equiv \frac{\omega_J}{2\pi}\int_0^{2\pi/\omega_J}\text{d}t\,S^+_{i\alpha,j\beta}(\Omega,t)  ,\\
\bar{S}^-_{i\alpha,j\beta}(\Omega) &\equiv \frac{\omega_J}{2\pi}\int_0^{2\pi/\omega_J}\text{d}t\,S^-_{i\alpha,j\beta}(\Omega,t)  .
\label{eq:Sminus}
\end{align}
It is clear from these expressions that emission and absorption noises are trivially related when flipping the sign of the probing frequency, namely $\bar{S}^+_{i\alpha,j\beta}(\Omega) = \bar{S}^-_{i\alpha,j\beta}(-\Omega)$. A theory for the detection of photo-assisted finite frequency shot noise has been presented in Ref. \onlinecite{prb81_chevallier} with an application to the fractional quantum Hall effect, and it is directly applicable to superconducting systems.

\begin{widetext}

Substituting the expression for the current correlator, Eq.\eqref{real_time_noise_RAK}, in terms of the dots Green's functions and leads self-energies back into Eqs.~\eqref{eq:Splus}-\eqref{eq:Sminus}, and focusing on the behavior at zero frequency, one obtains, after performing all 4 time integrals explicitly
\begin{align}
\bar{S}_{i\alpha,j\beta}(\Omega=0) &=
-\frac{(2\pi)^3}{2}{\rm{}Re}
\int_0^{\omega_J} \text{d}\omega  ~ \text{Tr}^{(NH)}\left\{ \sigma_{z}\left(\tilde{\Sigma}_{i}^{K}\tilde{G}^{A}+\tilde{\Sigma}_{i}^{R}\tilde{G}^{K}
-\tilde{\Sigma}_{i}^{A}\tilde{G}^{A}+\tilde{\Sigma}_{i}^{R}\tilde{G}^{R}\right)^{\alpha\beta}_{\omega}\right. \notag\\
&\hspace{15em}\times \sigma_{z}\left(\tilde{\Sigma}_{j}^{K}\tilde{G}^{A}+\tilde{\Sigma}_{j}^{R}\tilde{G}^{K}
+ \tilde{\Sigma}_{j}^{A}\tilde{G}^{A}-\tilde{\Sigma}_{j}^{R}\tilde{G}^{R}\right)^{\beta\alpha}_{\omega} \notag\\
&\hspace{12em}-\sigma_{z}\left(\tilde{\Sigma}_{i}^{R}\tilde{G}^{R}\tilde{\Sigma}_{j}^{K}+\tilde{\Sigma}_{i}^{K}\tilde{G}^{A}\tilde{\Sigma}_{j}^{A}+\tilde{\Sigma}_{i}^{R}\tilde{G}^{K}\tilde{\Sigma}_{j}^{A} -\tilde{\Sigma}_{i}^{A}\tilde{G}^{A}\tilde{\Sigma}_{j}^{A}+\tilde{\Sigma}_{i}^{R}\tilde{G}^{R}\tilde{\Sigma}_{j}^{R}\right)^{\alpha\beta}_{\omega}\notag\\
&\hspace{15em}\left. \times\sigma_{z}\left(\tilde{G}^{K}+\tilde{G}^{A}-\tilde{G}^{R}\right)^{\beta\alpha}_{\omega}\right\} \notag\\
&\hspace{0em}- \frac{2\pi \delta_{ij}}{4}  \mathrm{Re} \int_0^{\omega_J} \text{d}\omega ~\text{Tr}^{(NH)}\left\{ \sigma_z \left(\tilde{\Sigma}^A_{i} - \tilde{\Sigma}^R_{i} - \tilde{\Sigma}^K_{i}\right)^{\alpha \beta}_{\omega}  \sigma_z \left(\tilde{G}^A - \tilde{G}^R + \tilde{G}^K \right)^{ \beta \alpha}_{\omega}\right.   \notag\\ 
&\hspace{12em} + \left.\sigma_z \left(-\tilde{G}^A + \tilde{G}^R + \tilde{G}^K \right)^{\alpha \beta}_{\omega} \sigma_z \left(-\tilde{\Sigma}^A_{i} + \tilde{\Sigma}^R_{i} - \tilde{\Sigma}^K_{i}\right)^{ \beta \alpha}_{\omega} \right\}
  ,
\label{eqch3_corr_noise}
\end{align}
where we dropped the superscript $+/-$  for  $\bar{S}_{i\alpha,j\beta}$ since the two lead to the same result at zero frequency and we introduced $\text{Tr}^{(NH)}$ as the trace in Nambu-harmonics space.

It is also possible to obtain compact expressions when the noise is evaluated at harmonics $l \omega_J$ ($l$ integer) of the Josephson frequency
\begin{align}
\bar{S}^-_{i\alpha,j\beta}(l \omega_J) &=
-\frac{(2\pi)^3}{2}{\rm{}Re}
\sum_{n,p=-\infty}^{+\infty}\int_0^{\omega_J} d\omega \text{Tr}^{(N)}\Bigg\{\sigma_{z}\left(\tilde{\Sigma}_{i}^{K}\tilde{G}^{A}+\tilde{\Sigma}_{i}^{R}\tilde{G}^{K}
-\tilde{\Sigma}_{i}^{A}\tilde{G}^{A}+\tilde{\Sigma}_{i}^{R}\tilde{G}^{R}\right)^{\alpha,\beta; n,p+l}_{\omega}\notag\\
&\hspace{16em}\times \sigma_{z}\left(\tilde{\Sigma}_{j}^{K}\tilde{G}^{A}+\tilde{\Sigma}_{j}^{R}\tilde{G}^{K}
+\tilde{\Sigma}_{j}^{A}\tilde{G}^{A}-\tilde{\Sigma}_{j}^{R}\tilde{G}^{R}\right)^{\beta,\alpha; p, n-l}_{\omega}\notag\\
&\hspace{12em} -\sigma_{z}\left(\tilde{\Sigma}_{i}^{R}\tilde{G}^{R}\tilde{\Sigma}_{j}^{K}+\tilde{\Sigma}_{i}^{K}\tilde{G}^{A}\tilde{\Sigma}_{j}^{A}+\tilde{\Sigma}_{i}^{R}\tilde{G}^{K}\tilde{\Sigma}_{j}^{A}
-\tilde{\Sigma}_{i}^{A}\tilde{G}^{A}\tilde{\Sigma}_{j}^{A}+\tilde{\Sigma}_{i}^{R}\tilde{G}^{R}\tilde{\Sigma}_{j}^{R}\right)^{\alpha,\beta; n,p+l}_{\omega}\notag\\
&\hspace{16em} \times\sigma_{z}\left(\tilde{G}^{K}+\tilde{G}^{A}-\tilde{G}^{R}\right)^{\beta,\alpha; p,n-l}_{\omega}\Bigg\} \notag\\ 
&- \frac{2\pi \delta_{ij}}{4}  \mathrm{Re} \sum_{n,p=-\infty}^{+\infty} \int_0^{\omega_J} \text{d}\omega ~\text{Tr}^{(N)}\left\{ \sigma_z \left(\tilde{\Sigma}^A_{i} - \tilde{\Sigma}^R_{i} - \tilde{\Sigma}^K_{i}\right)^{\alpha, \beta; n,p+l}_{\omega}  \sigma_z \left(\tilde{G}^A - \tilde{G}^R + \tilde{G}^K \right)^{ \beta, \alpha; p,n-l}_{\omega}\right.   \notag\\ 
&\hspace{12em}+ \left.\sigma_z \left(-\tilde{G}^A + \tilde{G}^R + \tilde{G}^K \right)^{\alpha, \beta; p,n-l}_{\omega} \sigma_z \left(-\tilde{\Sigma}^A_{i} + \tilde{\Sigma}^R_{i} - \tilde{\Sigma}^K_{i}\right)^{ \beta, \alpha; p,n-l}_{\omega} \right\}
~.\label{eqch3_corr_noise_pf}
\end{align}

Eqs. (\ref{eqch3_corr_current}) and (\ref{eqch3_corr_noise_pf})  for the Josephson current and the noise deserve some attention. They are not specific to the dot geometry, lead configuration, or tunnel amplitudes  whatsoever. The self energies are known [see Eq.~\eqref{eq_ch3_model_self}], and the matrix elements of the dot Green's function need ``only'' to be solved via Dyson's equation, with a recursive scheme, or by a direct matrix inversion.

For the application of this study, our priority is to investigate both signals at zero frequency. Nevertheless, we stress out that in past works,\cite{prb80_jonckheere} the Josephson current harmonics have provided a useful test for the presence of decoherence effects. Indeed, in a superconducting junction with an embedded  quantum dot  side-coupled to a normal metal reservoir (a source of decoherence) non zero harmonics of the Josephson current are suppressed when the coupling to this normal metal reservoir is increased. Here, we point out that we can achieve a further characterization of the transport characteristics of the device via the noise harmonics, with the potential to incorporate the monitoring of decoherence effects. 

\end{widetext}

\subsection{Noise characteristics of the quartet resonance}\label{sec_nums}

To be more specific, we focus on the quartet resonance $V_a=-V_b=V>0$. Moreover, we adopt an antisymmetric position for the dots $\epsilon_a=-\epsilon_b=\epsilon>0$ to optimize CP splitting.~\cite{prb83_chevallier} We assume identical superconducting gap for all leads, $\Delta_j = \Delta$, and symmetric couplings between superconductors and dots $\Gamma_{j\alpha\beta}=\Gamma$ while forbidding interdot tunneling, $t_d=0$. The quartet phase $\varphi_Q=\varphi_a+\varphi_b-2\varphi_0$ is monitored through $\varphi_b=\varphi_Q$, the two others being set to zero, $\varphi_0=\varphi_a=0$. We present results for the DC currents in the leads $I_a\equiv {\cal I}_{aa}^{0}$ and $I_b\equiv{\cal I}_{bb}^0$ (cf. Eq.~\eqref{eqch3_corr_current}) as well as for the zero-frequency correlations of these currents $S_{aa}\equiv\bar{S}_{aa,aa}(\Omega=0)$, $S_{bb}\equiv\bar{S}_{bb,bb}(\Omega=0)$ and $S_{ab}\equiv\bar{S}_{aa,bb}(\Omega=0)$ (cf. Eq.~\eqref{eqch3_corr_noise}). $S_{aa}$ and $S_{bb}$ are referred to as noise autocorrelations whereas $S_{ab}$ are the noise crossed correlations. Fano factors for the lateral leads are given by the standard definition: 
\be
{\cal F}_j=\frac{S_{jj}}{2I_j}\quad\text{for}\quad j=a,b  .
\ee
Similarly, a Fano factor can be defined for the central lead from the sum of the currents, leading to meaningful information about noise crossed correlations
\be
{\cal F}_0=\frac{S_{aa}+S_{bb}+2S_{ab}}{2(I_a+I_b)}  .
\ee

\begin{figure}[tbp]
\centerline{\includegraphics[width=0.5\textwidth]{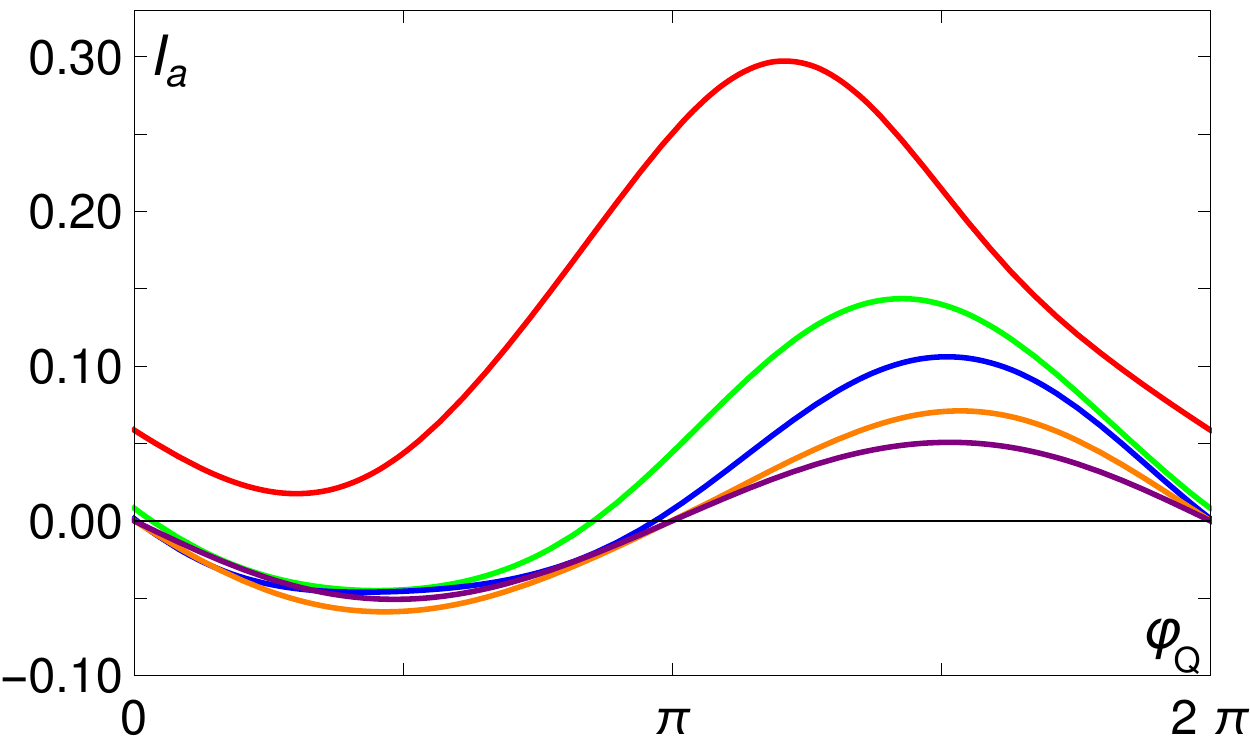}}
\caption{Current $I_a$ (in units of $ \Delta$) as a function of the phase $\varphi_Q$ in the QPC regime, for voltages $V=0.65 \Delta$, $0.55\Delta$, $0.45\Delta$, $0.35\Delta$, $0.17\Delta$ 
(in order of decreasing amplitude).
Current $I_b$ can be deduced from $I_b(-\varphi_Q) = -I_a(\varphi_Q)$.}\label{fig:IaQPC}
\end{figure}

The properties of the current in this system were studied by some of the authors in 
Ref.~[\onlinecite{prb87_jonckheere}].
It was shown that the current can be broken down into three qualitatively different components.
First the quasiparticle current $I^{qp}$, due to the multiple Andreev processes, which is independent of the quartet phase. It is an odd function of voltage, and thus contributes to the currents $I_a$ and $I_b$ with an opposite sign since the voltages of leads $a$ and $b$ satisfy $V_a = - V_b$. Then the ``phase-MAR'' component $I^{phMAR}$, which is due to interference effects between multiple Andreev reflection and phase dependent processes. While it is odd in voltage like the quasiparticle current, $I^{phMAR}$ is, as suggested by its name, phase-dependent, and an even function of the quartet phase. Finally, the multipair coherent current $I^{MP}$, which is carried by the exchange of multiple pairs between superconducting leads, is an odd function of the quartet phase. This component is however an even function of the voltages, and thus contributes equally to $I_a$ and $I_b$.

When the voltage $V$ becomes small enough,  the contributions $I^{qp}$ and $I^{phMAR}$ become small compared with the coherent multipair current, which thus dominates for a non-zero quartet phase. This explains why the two currents $I_a$ and $I_b$ become equal in the low-voltage regime, and the current phase relation $I_a(\varphi_Q)$ is that of a $\pi$ junction, which can be justified by the spin singlet nature of CP pairing. When the voltage is increased, the multiple Andreev processes become more important, and the two currents $I_a$ and $I_b$ start deviating from each other. Within the symmetries that result from the choice of parameters we have used, we always have $I_b(-\varphi_Q)=-I_a(\varphi_Q)$.

In what follows, we investigate two different regimes:
(i) the "quantum point contact" (QPC) regime which is obtained for dot energy levels placed outside the gap of the central superconductor $\epsilon>\Delta$ and for large couplings $\Gamma>\Delta$, which imply weak energy filtering from the quantum dots;
(ii) the resonant dots regime which is obtained for QD energy levels within the gap $\epsilon<\Delta$ and moderate couplings $\Gamma<\Delta$ which imply efficient energy filtering by the quantum dots.   

\begin{figure}[tbp]
\centering
\includegraphics[width=0.45\textwidth]{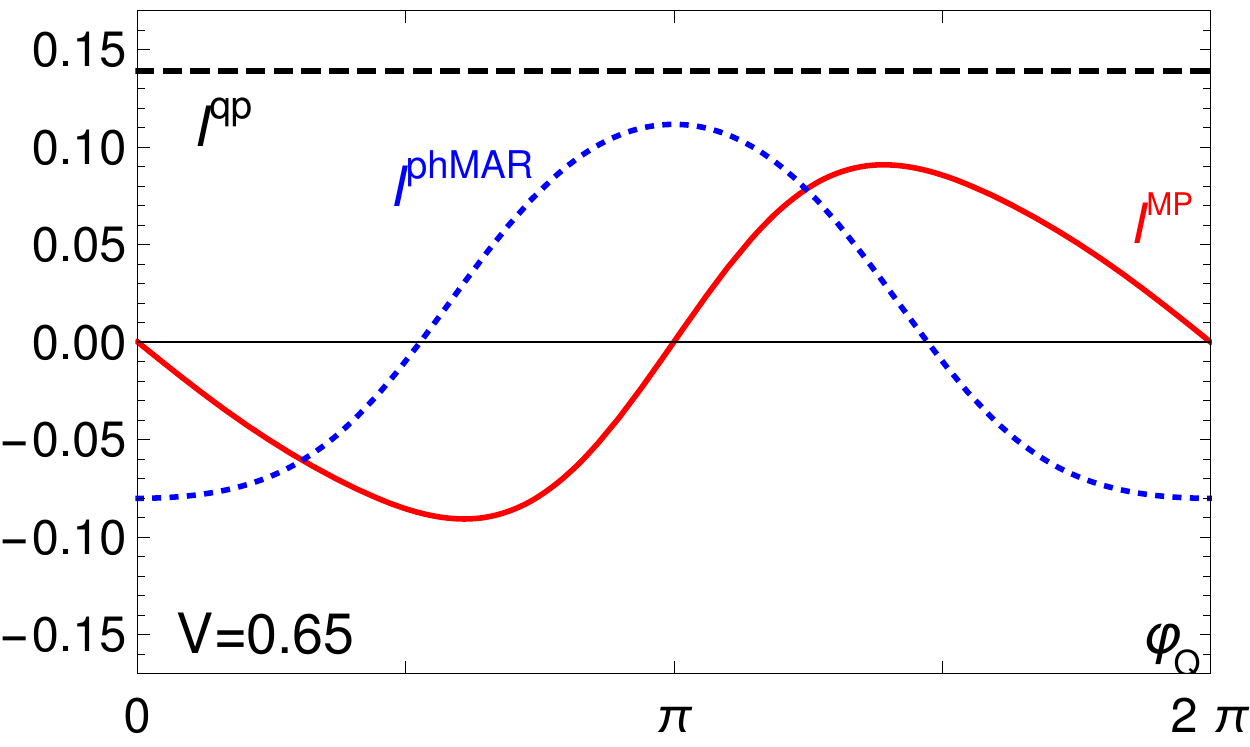}
\includegraphics[width=0.45\textwidth]{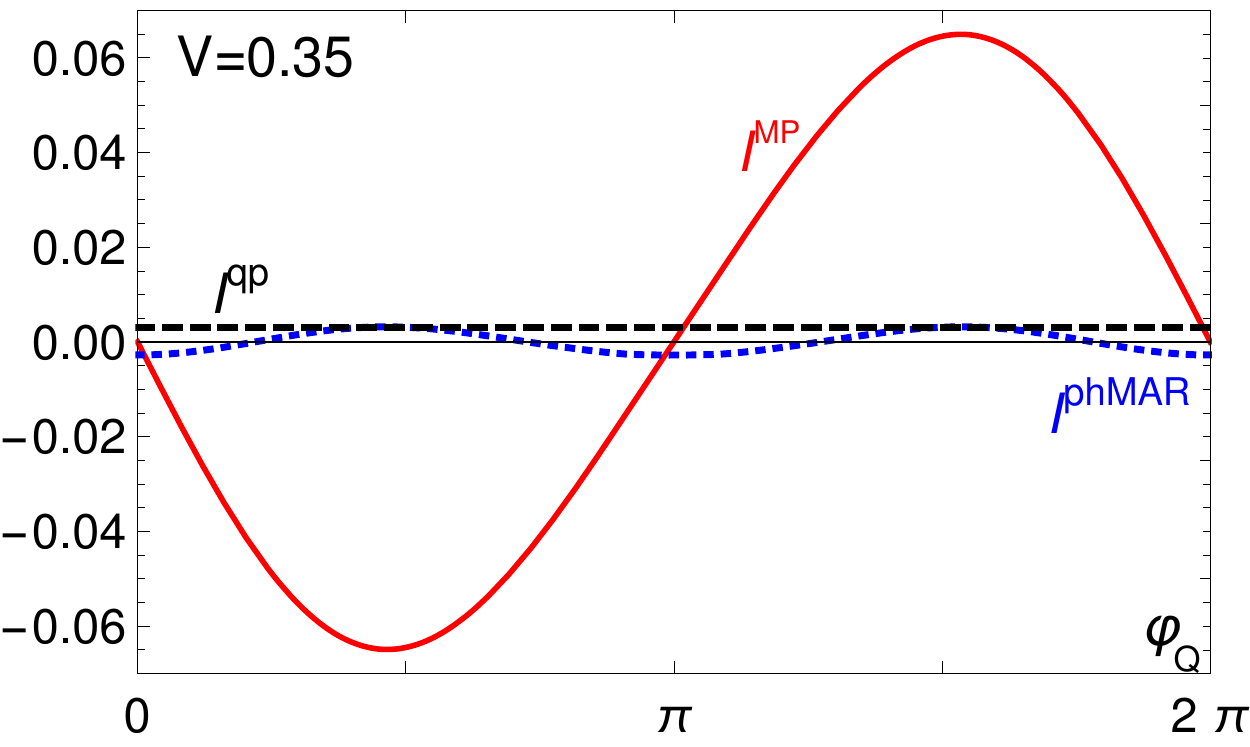}
\caption{Decomposition of the current $I_a$ in terms of the quasiparticle current $I^{qp}$, the
phase-MAR current $I^{phMAR}$ and the coherent multipair current $I^{MP}$ (all in units of $ \Delta$), as a function of the phase $\varphi_Q$ in the QPC regime,
for voltages $V=0.65\Delta$ (top) and $V=0.35\Delta$ (bottom).
For $V=0.65\Delta$, the three contributions are of the same order of magnitude. For $V=0.35\Delta$, the voltage
is low enough to have $I^{MP}$ strongly dominate over the other two contributions.}\label{fig:IadecompQPC}
\end{figure}

\subsubsection{QPC regime}

\begin{figure}[tbp]
\centering
\includegraphics[width=0.45\textwidth]{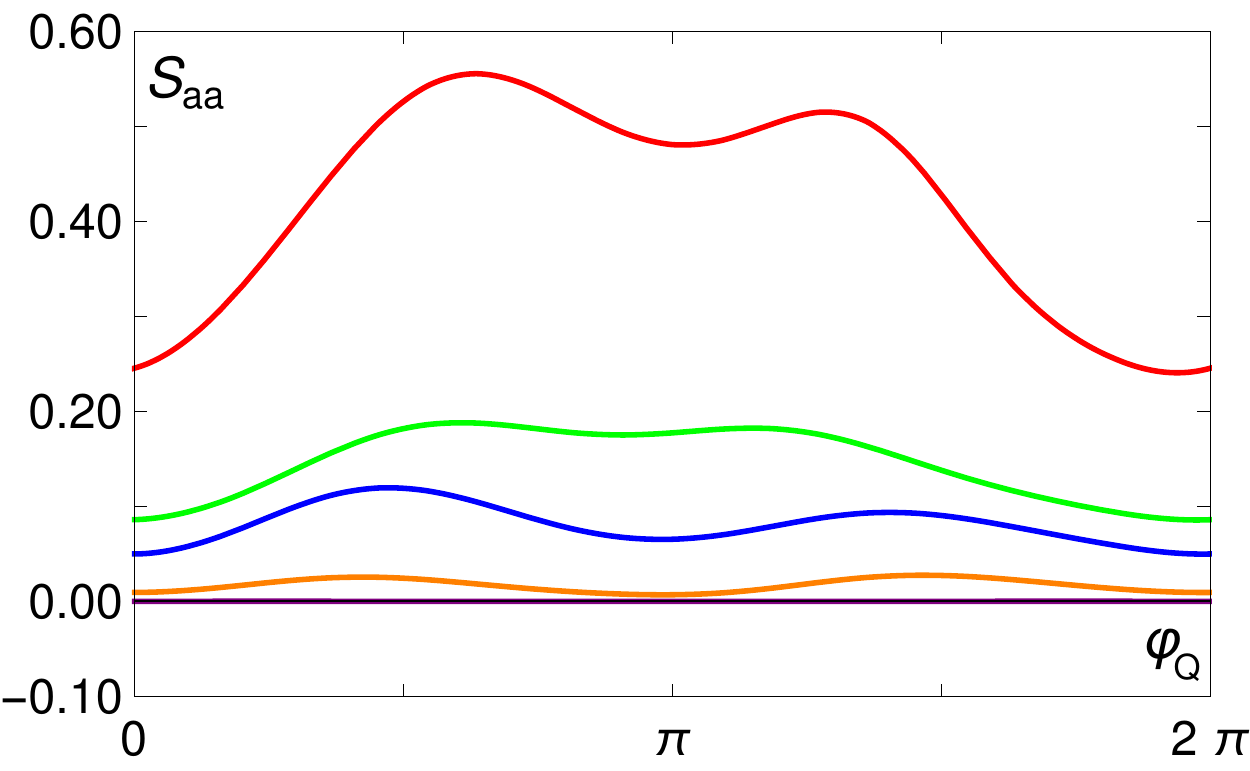}
\includegraphics[width=0.45\textwidth]{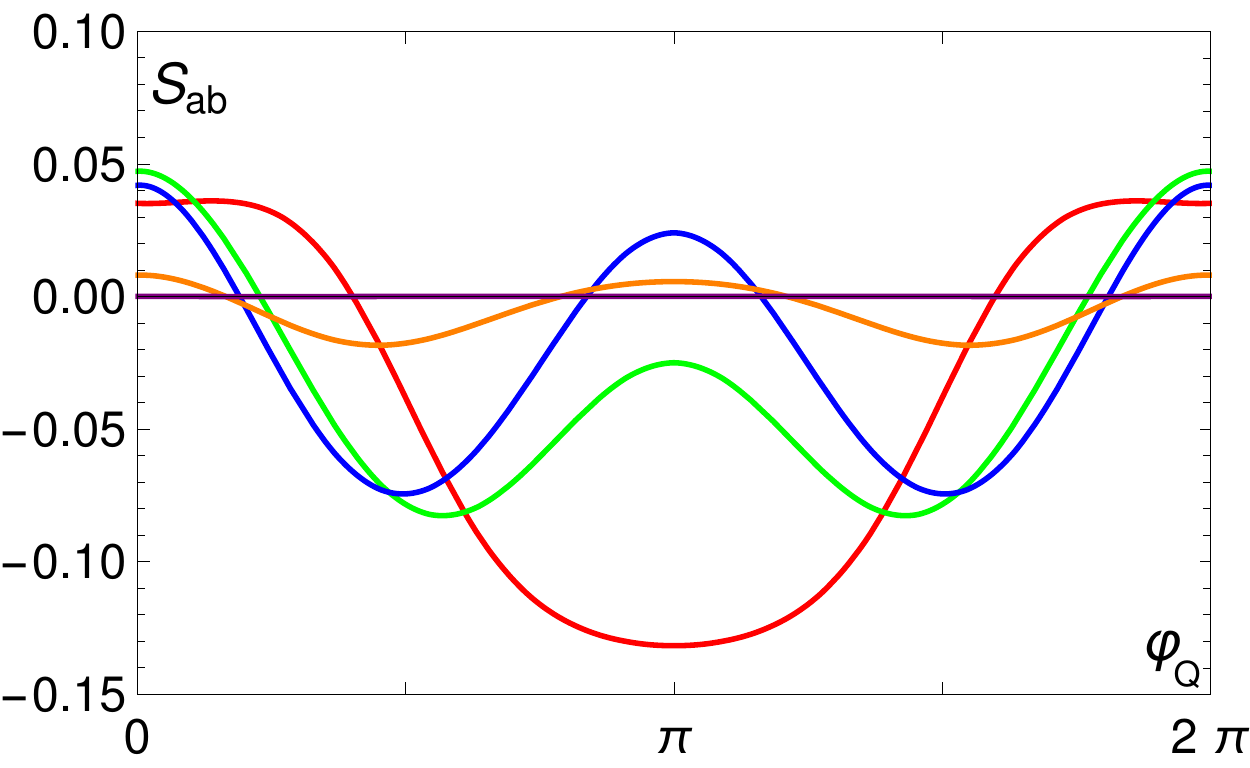}
\caption{Autocorrelation noise $S_{aa}$ (top) and crossed correlation noise $S_{ab}$ (bottom), given in units of $ \Delta$, as a function of the phase $\varphi_Q$ for voltages $V=0.65\Delta, 0.55\Delta, 0.45\Delta, 0.35\Delta, 0.17\Delta$ (in order of decreasing amplitude), in the QPC regime. Autocorrelations $S_{bb}$ can be deduced from $S_{bb}(\varphi_Q) = S_{aa}(-\varphi_Q)$.}\label{fig:SaaQPC}
\end{figure}

\begin{figure}[tbp]
\centering
\includegraphics[width=0.45\textwidth]{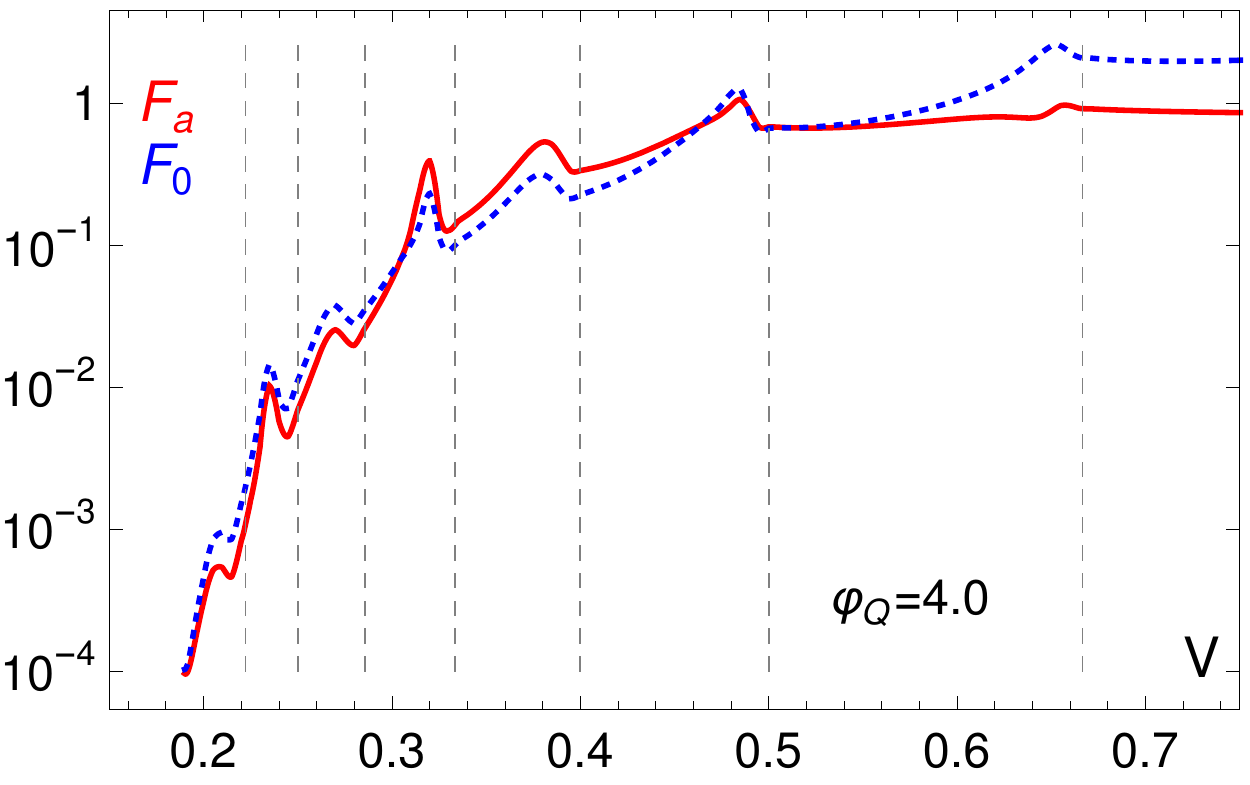}
\caption{Fano factors ${\cal F}_{a}$ (red, full line) and ${\cal F}_0$ (blue, dotted line) in the QPC regime, as a function of voltage, for $\varphi_Q=4$. Vertical lines indicate the location of the MAR onsets ($V=2 \Delta/n$).}\label{fig:FanoQPC}
\end{figure}

\begin{figure}
\includegraphics[width=0.35\textwidth]{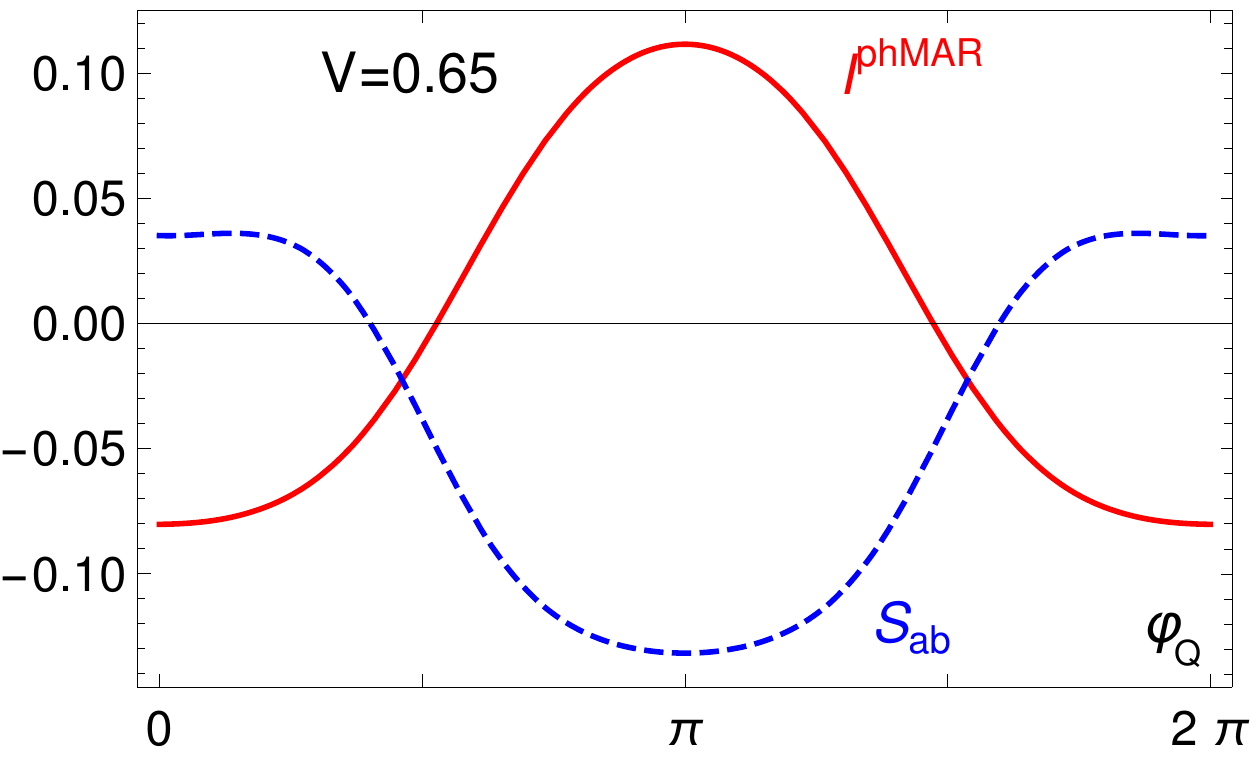} 
\includegraphics[width=0.36\textwidth]{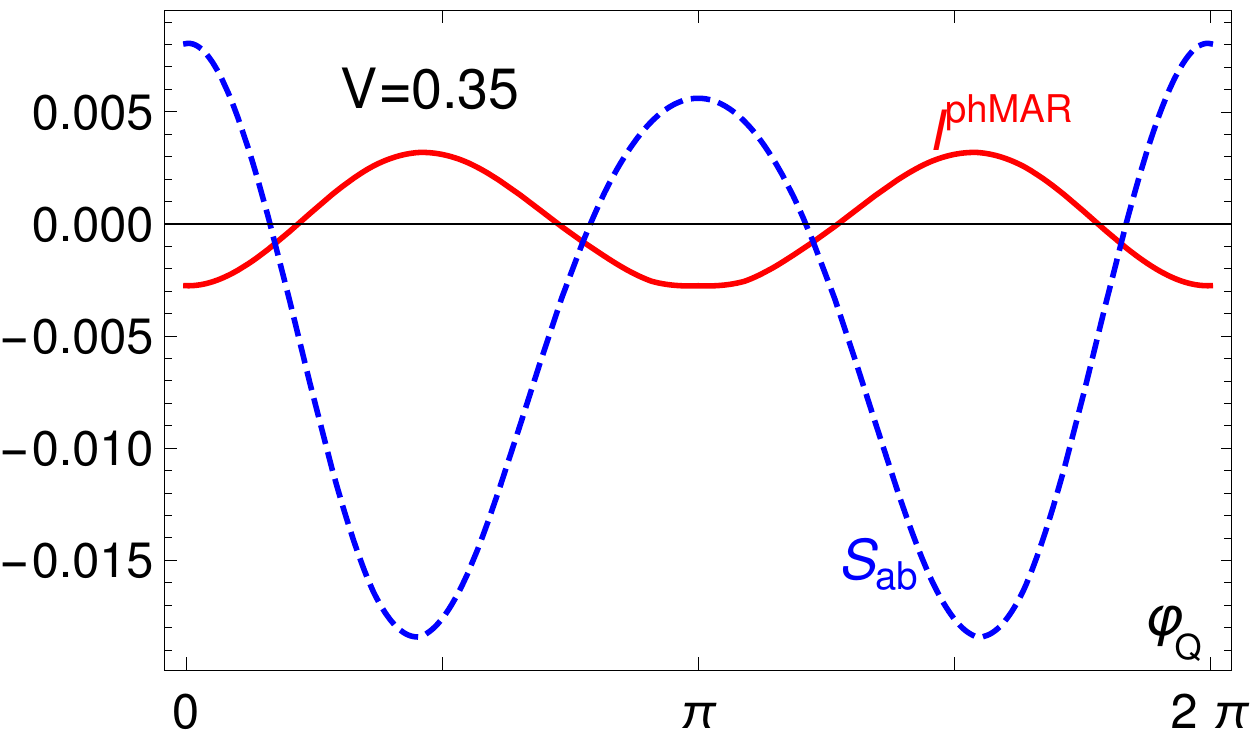}
\caption{Current component $I^{phMAR}$ (red, full curve), and noise crossed correlations $S_{ab}$ (blue, dashed curve),  in units of $ \Delta$, as a function of the phase $\varphi_Q$, for voltages $V=0.65 \Delta$ (top) and $V=0.35 \Delta$ (bottom). }
\label{fig:noisephMarQPC}
\end{figure}

\begin{figure}
\centerline{\includegraphics[width=0.45\textwidth]{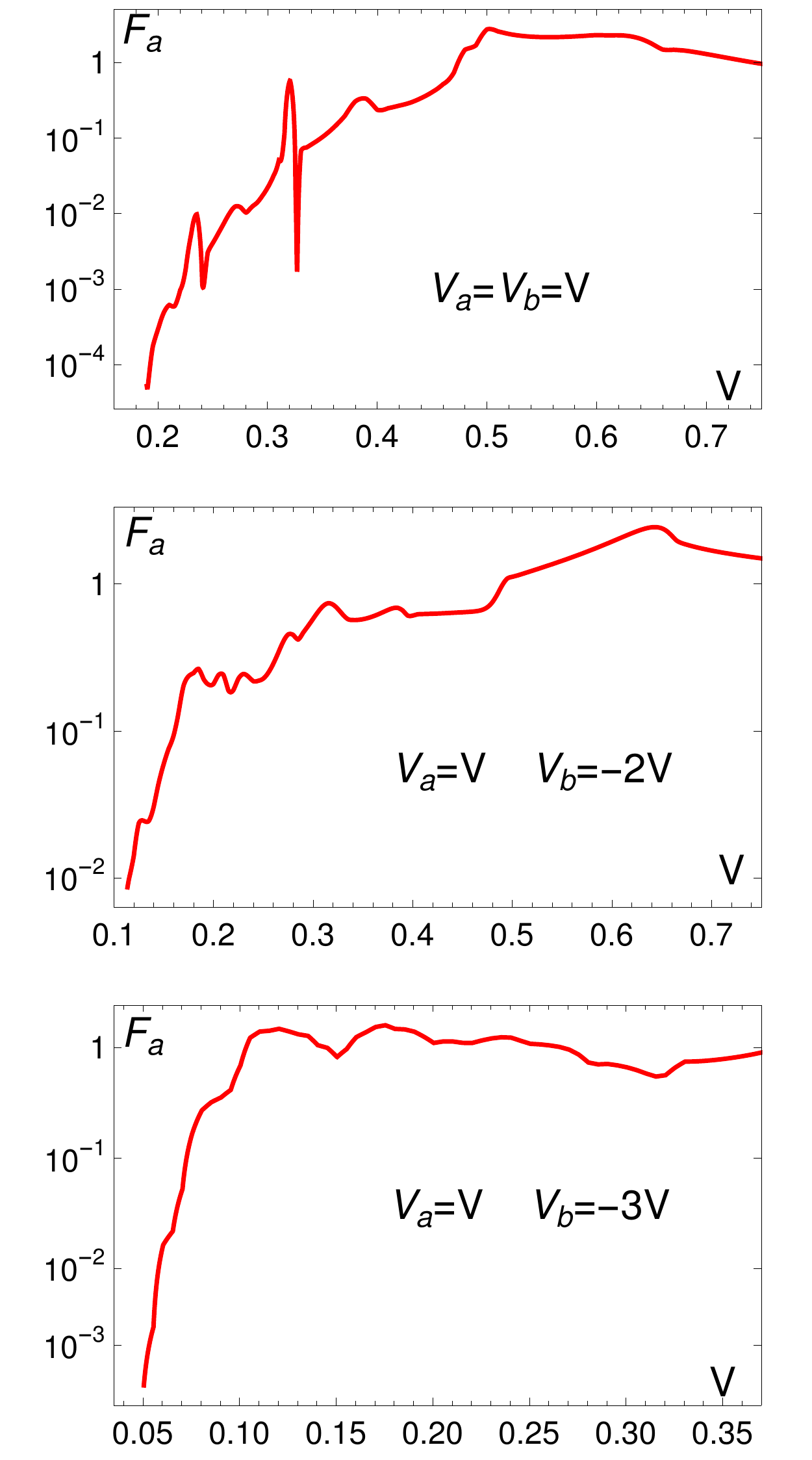}}
\caption{Fano factor ${\cal F}_a$ in the QPC regime, as a function of voltage, with $\varphi_a=0$ and $\varphi_b=4$, for three different voltages configurations.}
\label{fig:FanoMP}
\end{figure}

When the quantum dot energy levels are placed outside the gap of the central superconductor and the effective linewidths of the quantum dots are large compared with this gap, the two S-dot-S junctions behave like adjustable tunnel barriers, which justifies the nomenclature ``quantum point contact regime". This regime constitutes a natural starting point of our investigation as it provides an interesting insight into the physics at play in more complex situations. The results shown here in Figs.~\ref{fig:IaQPC}-\ref{fig:FanoMP} were obtained for the choice of parameters $\epsilon=6\Delta$ and $\Gamma=4\Delta$. 

The current $I_a$, in units of $\Delta$, as a function of the phase $\varphi_Q$ is shown in Fig.~\ref{fig:IaQPC} for
different values of the voltage $V$ [the current $I_b$ can be deduced from $I_b(-\varphi_Q) = - I_a(\varphi_Q)$]. For large values of the voltage (the largest here being $V=0.65 \Delta$), the current $I_a$ is positive for all phases, with a non-symmetric phase dependence. As the voltage is lowered, the current decreases in amplitude and tends towards a sinusoidal $\sim \sin(\varphi_Q+\pi)$ dependence. This change of behavior can be understood by decomposing the current into its three contributions:\cite{prb87_jonckheere}
\begin{align}
I^{qp} =& \frac{1}{2\pi}\int_0^{2\pi} \!\!\!\! d\varphi_Q \; I(\varphi_Q) \\
I^{phMAR} =& \frac{1}{2} \left[ I(\varphi_Q) + I(-\varphi_Q) \right] - I^{qp} \\
I^{MP} =& \frac{1}{2}  \left[ I(\varphi_Q) - I(-\varphi_Q) \right]   ,
\end{align}
where we recall that $I^{qp}$ is the quasiparticle current due to multiple Andreev processes, $I^{phMAR}$ is the phase-MAR current due to the interference between MAR processes and phase dependent processes, and $I^{MP}$ is the coherent multipair current associated with multiple Cooper pair resonance. 
Fig.~\ref{fig:IadecompQPC} shows the decomposition of the current $I_a$ for two different values of the applied voltage: $V=0.65 \Delta$ and $V=0.35\Delta$. For the larger applied voltage, the three components are roughly of the same magnitude, which explains the overall positive current, and its non-symmetric phase dependence. However, when $V$ is lowered, the $I^{qp}$ and $I^{phMAR}$ contributions decrease rapidly to zero, while the $I^{MP}$ component converges to a nonzero value, as shown for $V=0.35 \Delta$. Eventually, for low enough voltage, the current is fully due to the multipair component, which tends to a $\sim \sin(\varphi_Q + \pi)$ behavior. By the very nature of these multipair processes (which involve the same number of electrons being transferred from $S_0$ to $S_a$, and from $S_0$ to $S_b$) the currents $I_a$ and $I_b$ are simply equal. 

The noise as a function of the quartet phase is shown in Fig.~\ref{fig:SaaQPC} for different voltages. The autocorrelation noise $S_{aa}$ is naturally positive, and its amplitude decreases rapidly as the voltage is lowered. The crossed correlation noise $S_{ab}$ is mainly negative except for $\varphi_Q$ close to 0 (and $\pi$ at low voltage), and its amplitude shows a similar decrease as a function of voltage.
Note that the amplitude of the crossed correlations is smaller than that of the autocorrelations, a phenomenon which can be attributed to the local quasiparticle currents. Indeed, the local quasiparticle component of $I_a$, due to local MAR processes between electrodes $a$ and $0$, yields a finite contribution to the autocorrelations $S_{aa}$, while in the crossed correlations $S_{ab}$, the local quasiparticle current in $I_a$ and the one in $I_b$ are uncorrelated and thus do not contribute.

Comparing the results of Figs.~\ref{fig:IaQPC} and \ref{fig:SaaQPC}, it is interesting to point out that as one lowers the external voltage bias, the current assumes a finite value (for non-zero phase $\varphi_Q$) while the noise correlations tend to zero, so that the DC current becomes effectively noiseless at small voltage, a feature generally attributed to Josephson physics in standard two-terminal superconducting devices at equilibrium. 
Indeed, as argued above when studying the current, the low-bias transport properties rely uniquely on the exchange of CPs rather than dissipative processes (such as MAR or quasiparticle tunneling) which involve the continuum spectrum of the superconductors, thus explaining the strong similarities with a conventional Josephson junction. 
 This is better illustrated in Fig.~\ref{fig:FanoQPC} which shows the Fano factors ${\cal F}_a$ (for current $I_a$) and ${\cal F}_0$
(for the total current $I_a$ + $I_b$) as a function of the voltage bias for a generic phase $\varphi_Q=4$. The logarithmic scale clearly shows 
the rapid decrease of the Fano factor as $V \to 0$. Interestingly, the results of Fig.~\ref{fig:FanoQPC} also indicate the existence of peaks in the Fano factors near the MAR onsets $V=2\Delta/n$ (vertical dashed lines). While the presence of such structures is reminiscent of what is observed in a biased two-terminal junction separated by a QPC, the overall voltage-dependence is in sharp contrast with this situation as instead of vanishing, the Fano factor of such a device increases dramatically as  $1+  \mbox{Int}(2\Delta/V)$ at low voltages for low transparency,~\cite{prl82_cuevas} reflecting the increase in the transmitted charge through MAR processes. The observed drop of the Fano factor in our 3-superconductor system thus constitutes direct evidence that the DC signal of quartets is noiseless when lowering the voltage.

Further understanding of the noise behavior as a function of voltage can be obtained using the current decomposition in terms of quasiparticle, phase-MAR and multipair contributions. As argued above, in the low-voltage limit, the noise (both autocorrelations and crossed correlations) goes to zero while the current reduces to its coherent multipair component $I^{MP}$. From this, it is clear that the multipair component of the current is noiseless. 
As a result, we can thus expect that the small noise contribution at small to moderate voltage is directly related to the other two components of the current.  
This is confirmed in Fig.~\ref{fig:noisephMarQPC}, which shows the crossed correlations $S_{ab}$ together with the phase-MAR component $I^{phMAR}$ of the current, for $V=0.65 \Delta$ (top) and $V=0.35 \Delta$ (bottom). Quite strikingly, the phase dependence of the $I^{phMAR}$ and of $S_{ab}$ are directly correlated, which shows that the crossed correlations are indeed associated with $I^{phMAR}$ at least as far as the phase dependence goes (since $I^{qp}$ is phase independent). One can also see that the amplitudes of the current component and of the crossed correlations are somewhat similar. Crossed correlations decrease slower than $I^{phMAR}$, an effect that may be attributed to the quasiparticle current $I^{qp}$, which does contribute to the crossed correlations (through processes involving all three superconducting leads).

Finally, the significant drop of the Fano factor at very low voltage, which shows that the multipair process is noiseless, is not restricted to the quartet configuration. This is illustrated in Fig.~\ref{fig:FanoMP}, which presents the Fano factor  ${\cal F}_a$ as a function of the voltage bias, for three different voltage configurations:
$V_b = V_a$, $V_b = -2V_a$ and $V_b = -3V_a$. For each plot, the choice of phase is $\varphi_a=0$ and $\varphi_b=4$ (this choice does not impact qualitatively the results). As in Fig.~\ref{fig:FanoQPC} obtained for the quartet configuration, results are shown on a logarithmic scale, and one clearly sees that the Fano factor is strongly reduced at low enough voltage. As noticed for the quartet configuration, this is due to the existence of a finite DC current as $V \to 0$, while the noise vanishes for $V \to 0$.

\subsubsection{Resonant dots regime}

We now present results obtained for the set of parameters $\epsilon=0.6\Delta$ and $\Gamma=0.3\Delta$ which constitutes a regime of efficient energy filtering from the dots in the gap of the central superconductor, albeit still allowing multiple Andreev processes to be identified. 
The main qualitative difference with the QPC regime, which is common for all the investigations led in the resonant dots regime, is the presence of phase sensitive and voltage sensitive giant Fano factors even at low voltage when the Josephson current is fully ascribed to quartet processes. 

\begin{figure*}[t]
\begin{tabular}{cc}
\includegraphics[width=0.45\textwidth]{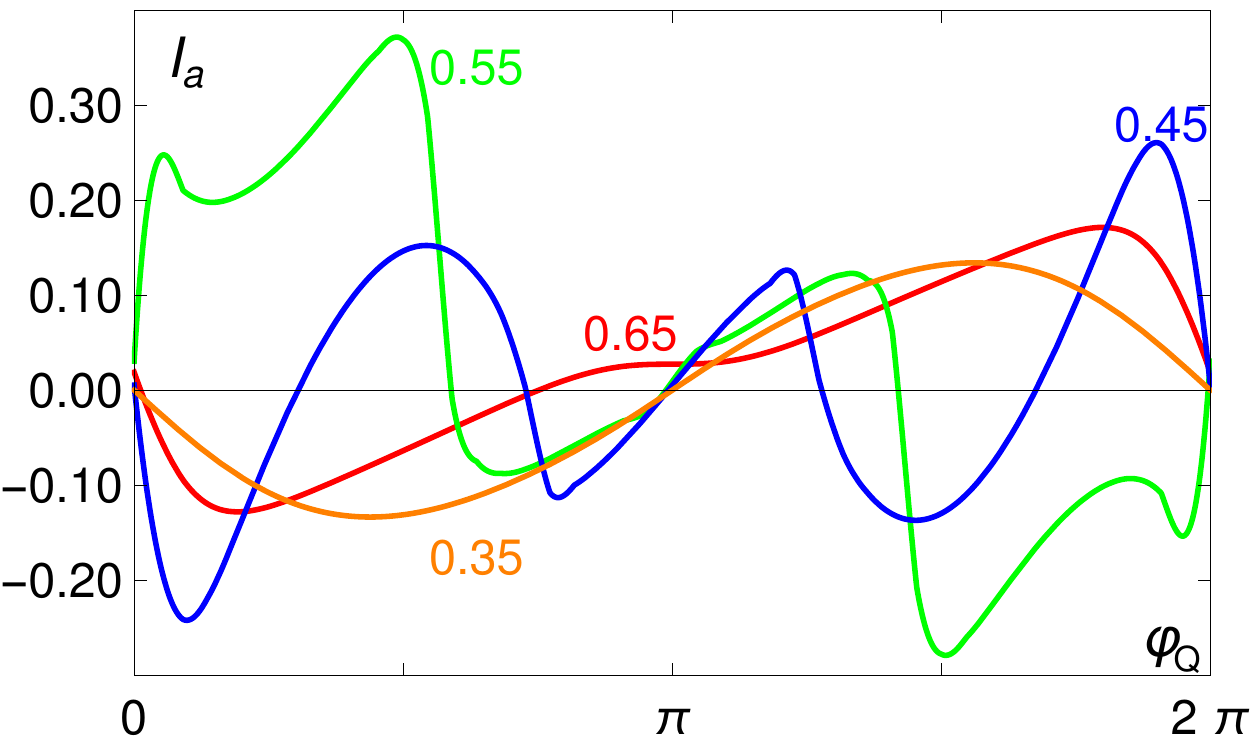} &
 \includegraphics[width=0.45\textwidth]{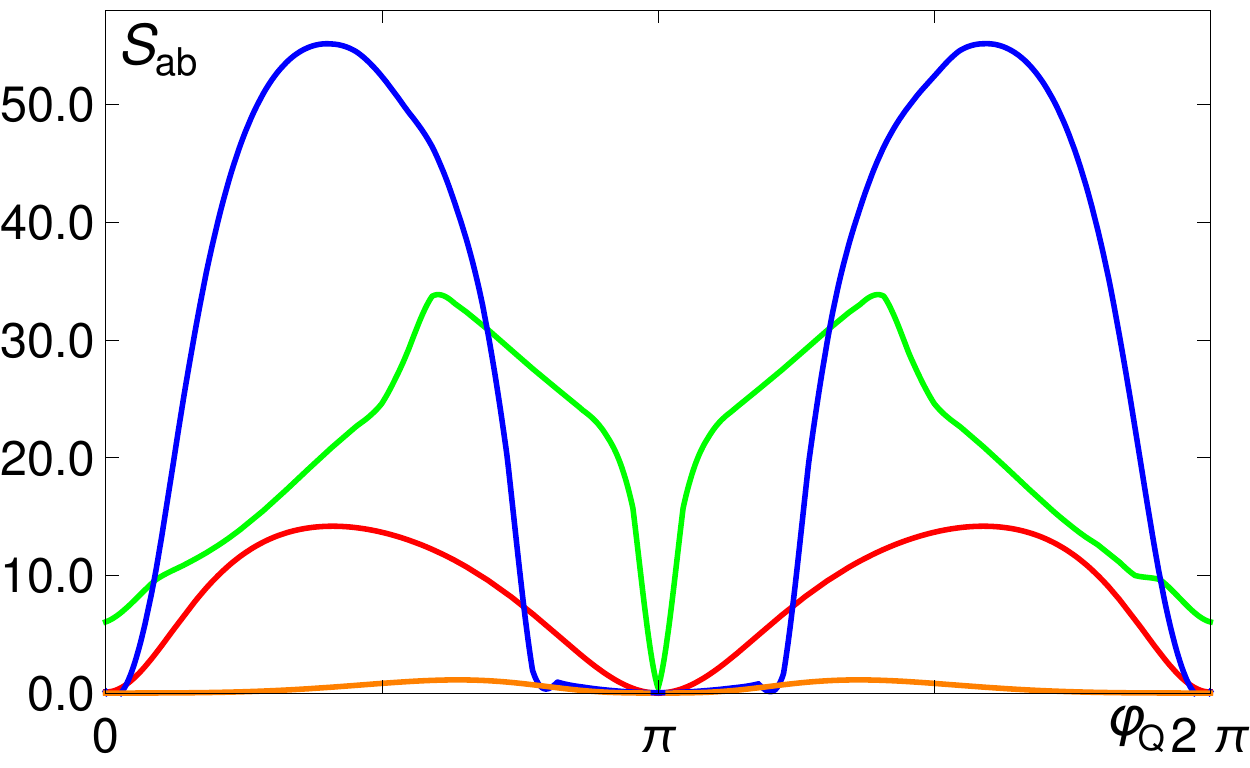} \\
\includegraphics[width=0.45\textwidth]{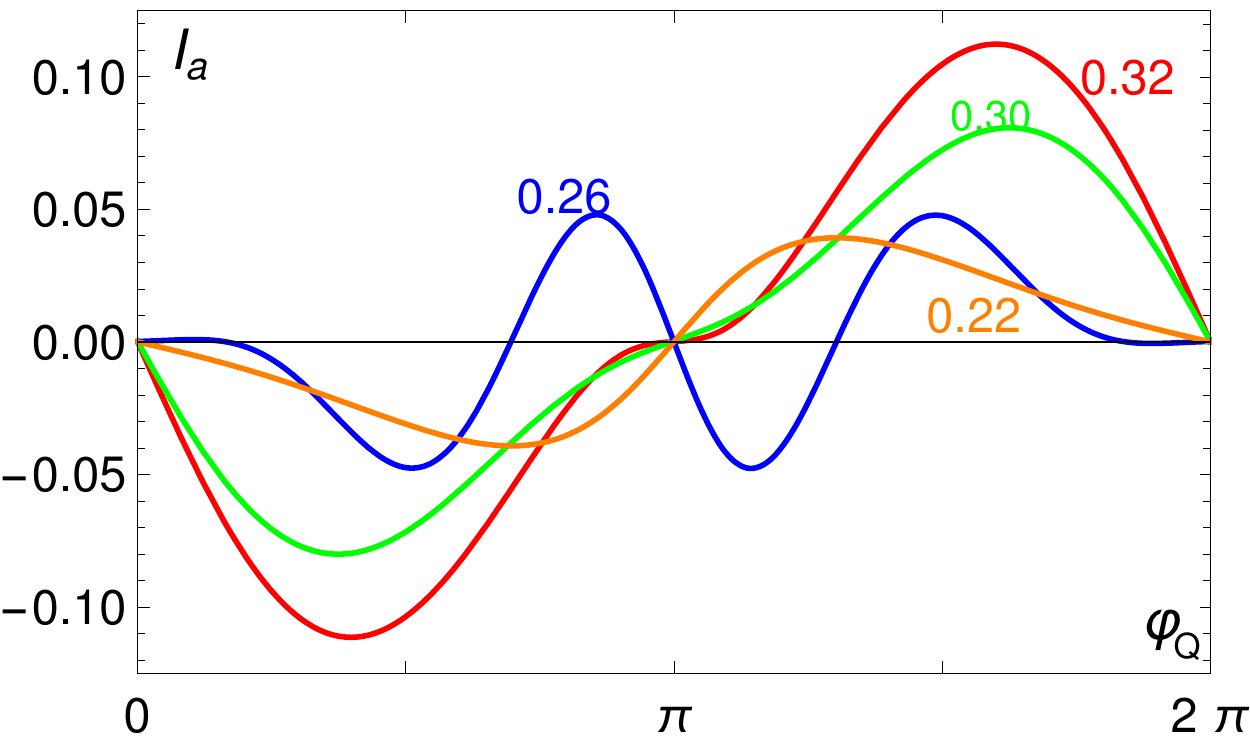}& 
 \includegraphics[width=0.45\textwidth]{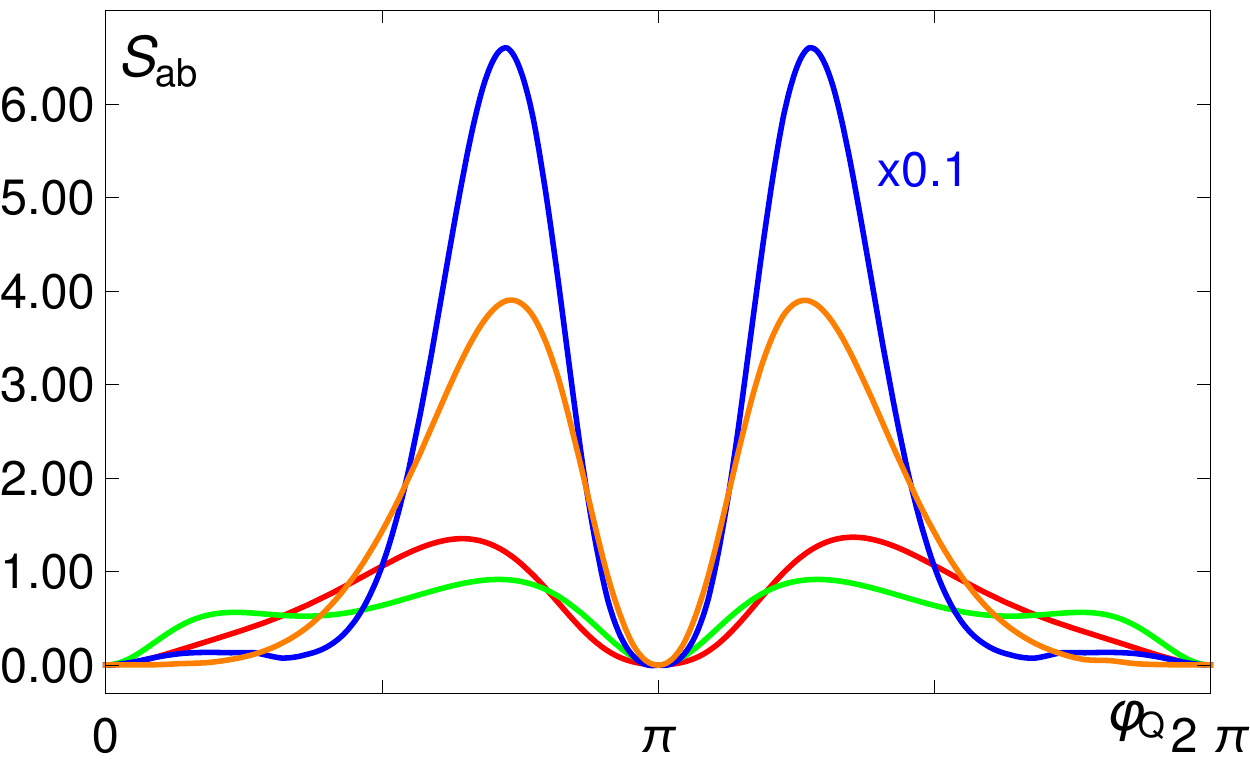}\\
\includegraphics[width=0.45\textwidth]{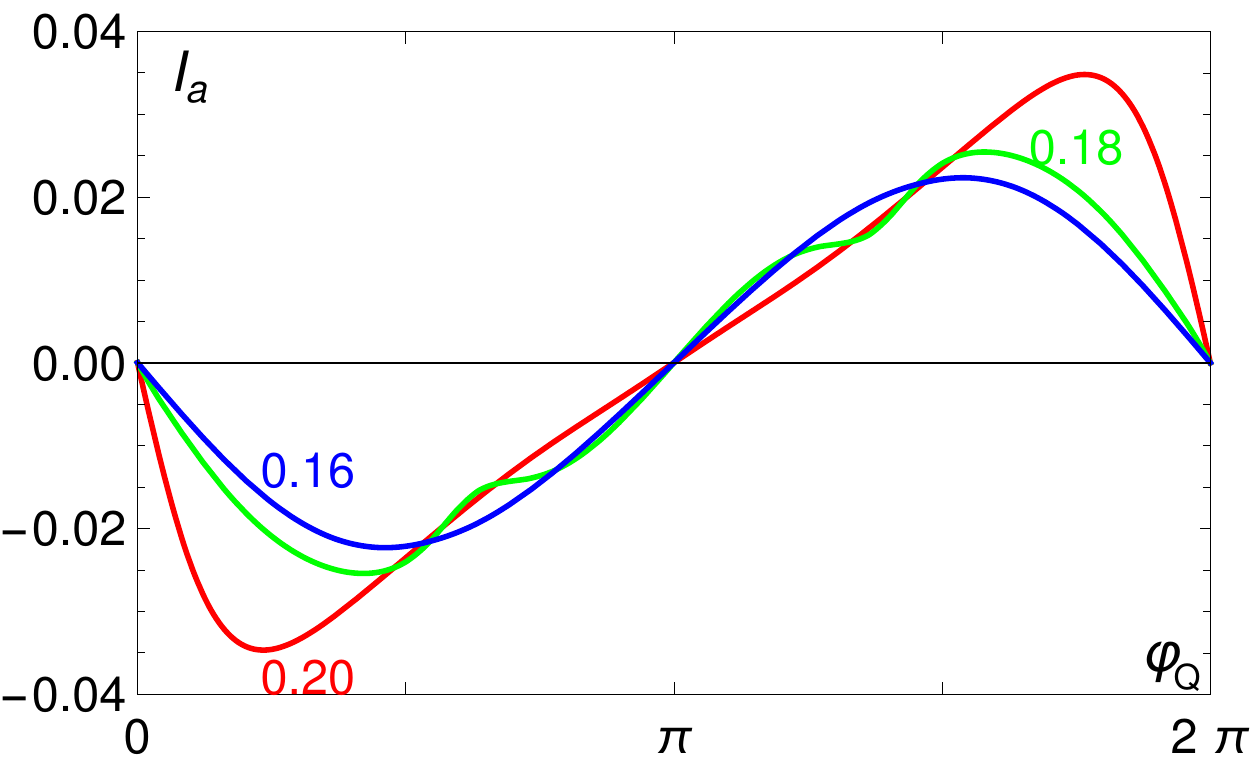}&
 \includegraphics[width=0.45\textwidth]{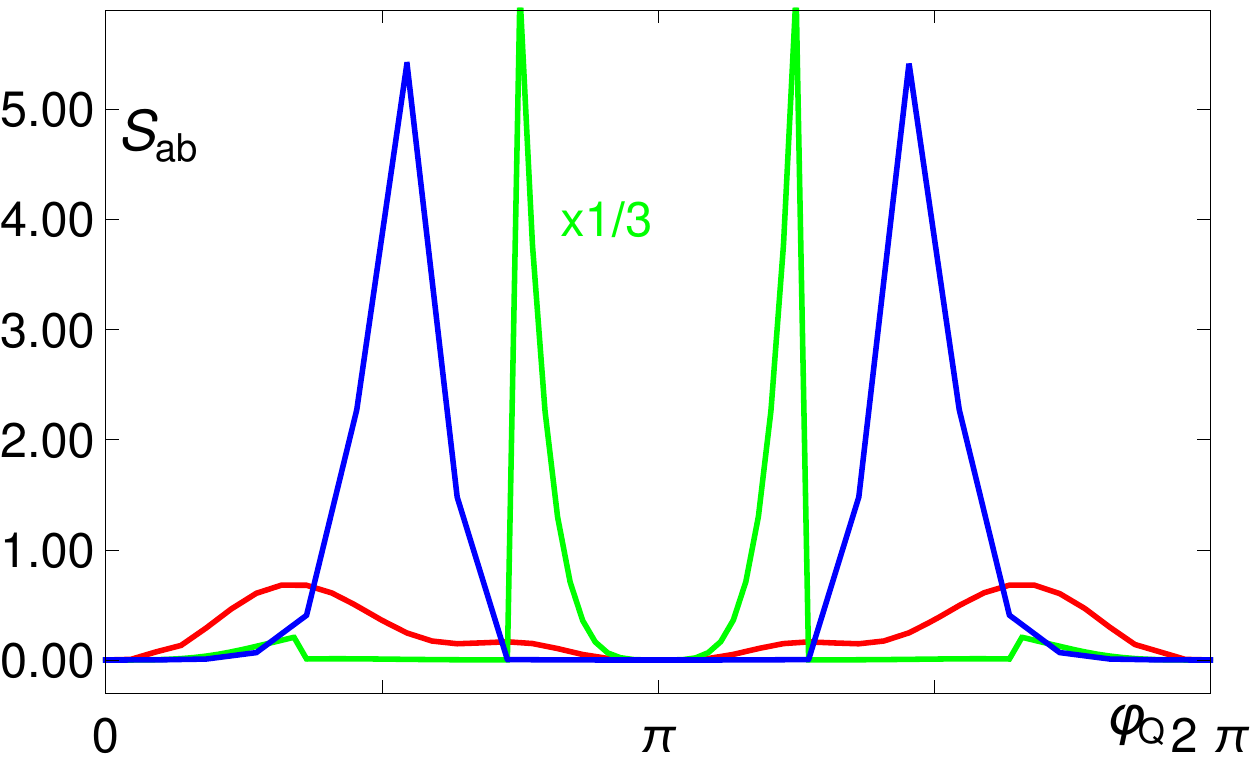}
\end{tabular}
\caption{Current and noise correlations (in units of $\Delta$) as a function of the phase $\varphi_Q$ for different values of voltage (from $V=0.65$ to $V=0.16$) in the resonant dots regime. The left column shows the current $I_a$, with the value of the voltage noted near each curve. The right column shows the crossed correlations $S_{ab}$ for the same values of voltages. Two curves for the $S_{ab}$ have been scaled to fit inside the plot: the one for $V=0.26$ (middle plot, factor 0.1) and the one for $V=0.18$ (bottom plot, factor $1/3$). }\label{ch3fig_hvolt_resonant}
\end{figure*}

Fig.~\ref{ch3fig_hvolt_resonant} shows the phase dependence of the current $I_a$ and the noise crossed correlations $S_{ab}$ for different voltages between $V=0.65\Delta$ and $V=0.16\Delta$. In all generality, the current $I_b(\varphi_Q)$ can be readily deduced from $I_b(-\varphi_Q)=-I_a(\varphi_Q)$, but except for the largest voltage considered, our results show that $I_b(\varphi_Q)\simeq I_a(\varphi_Q)$, which means that the current is largely dominated by quartet processes. 
Similarly, the noises $S_{aa}(\varphi_Q)$, $S_{bb}(\varphi_Q)$ and $S_{ab}(\varphi_Q)$ are nearly equal, except again for $V=0.65\Delta$ where there is a small but notable difference (which further increases for larger values of the voltage). In particular, this implies that the crossed correlation noise is positive for most of the voltage range, which constitutes yet another indication of the dominant character of quartet processes. This property of the currents and noises ultimately justifies our concentrating on $I_a$ and $S_{ab}$ in this regime.

Focusing on the results for the current (left column of Fig.~\ref{ch3fig_hvolt_resonant}), we see that for large and intermediate voltages (top and middle plots), the current-phase relation is quite complex and clearly non-harmonic. This is most marked around the value $V=0.60\Delta$ which is the voltage at which the dot level positions coincide with the chemical potentials of the left and right electrodes. 
As the voltage is lowered, the amplitude of the current oscillations decreases,  ultimately reaching
a $\pi$-shifted sinusoidal relation for low-enough voltage (bottom plot), typical of a pure quartet process far from resonance.

The behavior of the noise $S_{ab}$ in the resonant dots case (right column of Fig.~\ref{ch3fig_hvolt_resonant}) is qualitatively quite different from its counterpart from the QPC regime. We observe two striking characteristics.
First, the noise shows a highly sensitive phase dependence, with marked peaks at values of the phase which depend on the voltage bias. 
Secondly, the amplitude of the noise varies quite strongly and non-monotonically with voltage, reaching huge values, order of magnitudes larger than the corresponding maximum current at the same voltage.
This last property, combined with the decrease in amplitude of the current as one lowers the external voltage bias, leads to a giant Fano factor in the low-voltage regime.
This behavior is in stark contrast with the one described in the previous section for the QPC regime: while in both cases the current is largely dominated by quartet processes, here it is accompanied by large current fluctuations, which are thus enhanced by the resonant nature of the quantum dots.

\begin{figure}
\centerline{\includegraphics[width=8.cm]{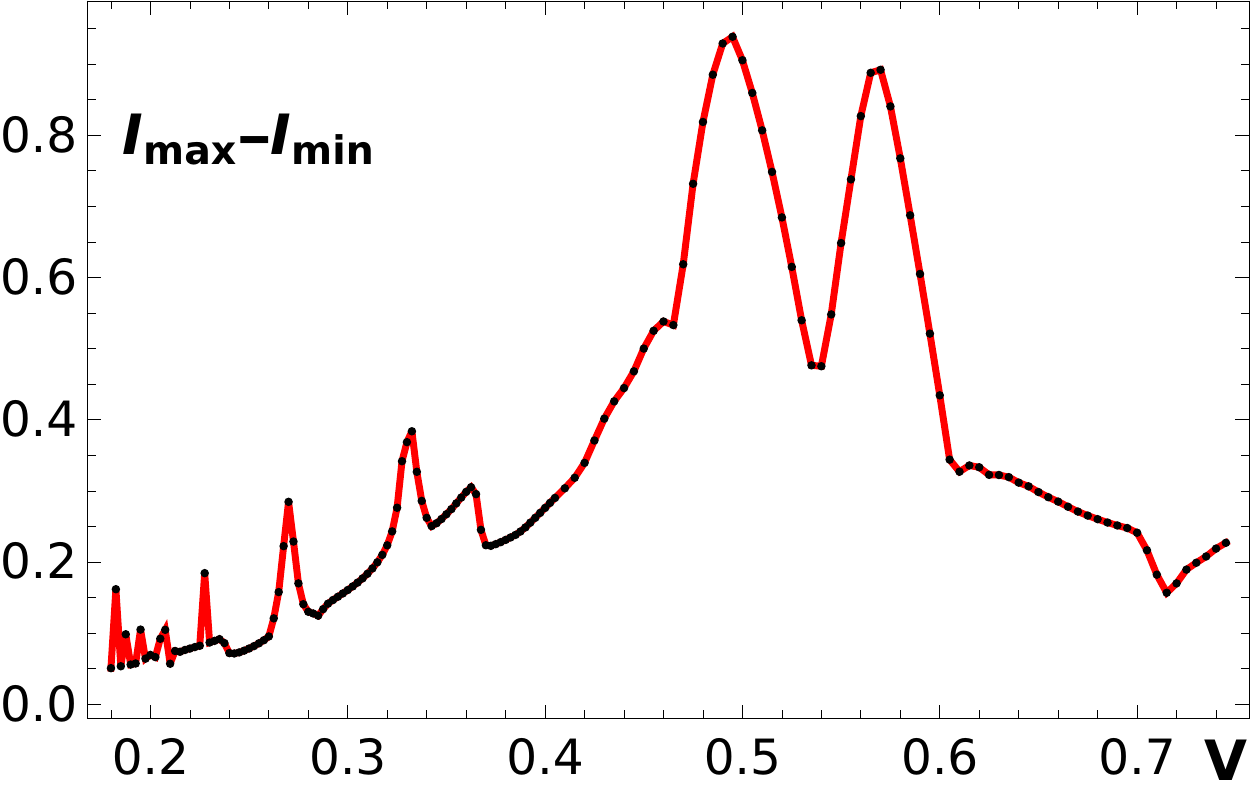}}
\caption{Amplitude of the current $I_a(\varphi_Q)$  (in units of $\Delta$), defined in Eq.~(\ref{eq:Iamp}), as a function of the voltage $V$, in the resonant dot regime ($\epsilon=0.6 \Delta$ , $\Gamma=0.3 \Delta$). The current is maximum near the voltage where the dot levels coincide with the lateral electrode chemical potentials ($V=\epsilon$).}
\label{fig:Iamp}
\end{figure}

\begin{figure}
\centerline{\includegraphics[width=8.cm]{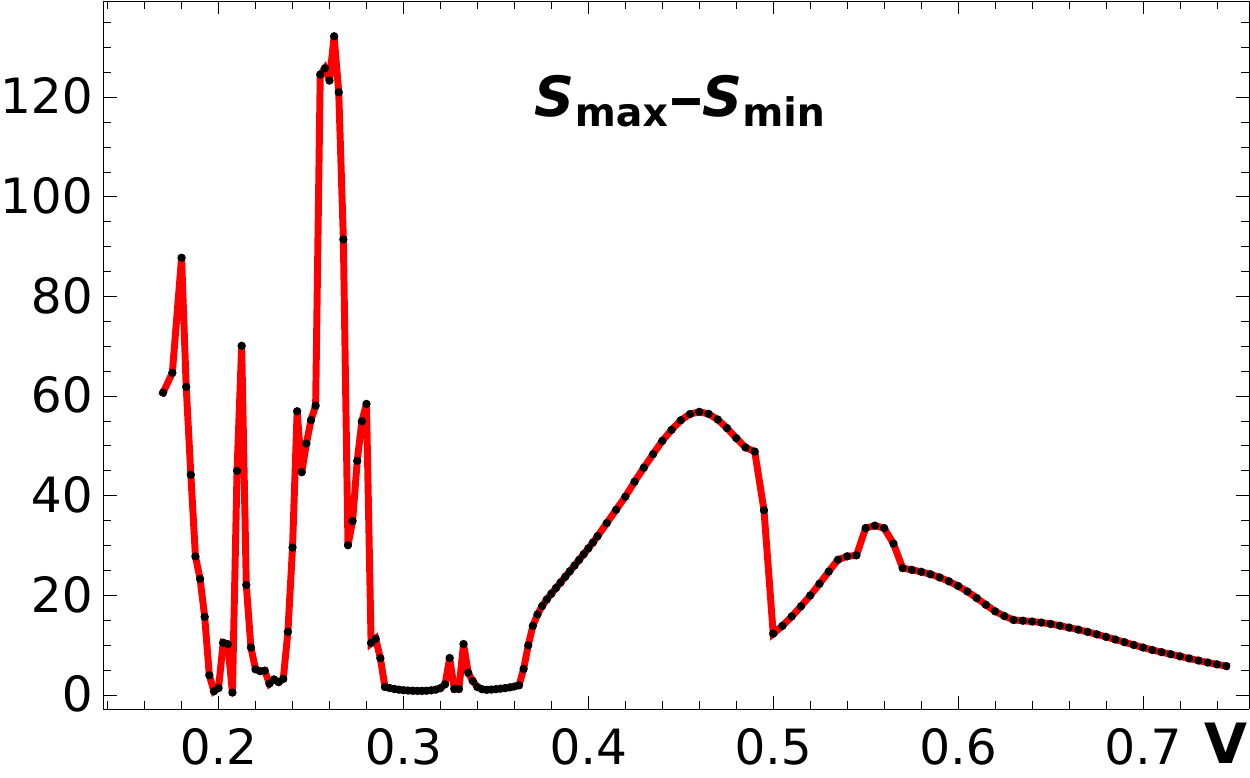}}
\caption{Amplitude of the crossed correlation noise $S_{ab}(\varphi_Q)$  (in units of $\Delta$), defined in Eq.~(\ref{eq:Samp}), as a function of the voltage $V$, in the resonant dot regime ($\epsilon=0.6 \Delta$ , $\Gamma=0.3 \Delta$). The noise shows strong variations at small voltages, with maximum values which are several orders of magnitude larger than the current.}
\label{fig:Samp}
\end{figure}

It is instructive to consider the overall amplitude of both current and noise, leaving aside the details of the phase dependence, as phase dependent measurements typically require a squid geometry, and we aim at computing quantities which are more directly accessible. To this aim, we consider the current amplitude at a given voltage defined as
\begin{equation}
I_\text{max} - I_\text{min} = \mbox{max}_{\varphi_Q} \left[I_a(\varphi_Q)\right] - \mbox{min}_{\varphi_Q} \left[ I_a(\varphi_Q) \right]   ,
\label{eq:Iamp}
\end{equation}
and similarly for the noise
\begin{equation}
S_\text{max} - S_\text{min} = \mbox{max}_{\varphi_Q} \left[ S_{ab}(\varphi_Q) \right] - \mbox{min}_{\varphi_Q} \left[ S_{ab}(\varphi_Q) \right]   .
\label{eq:Samp}
\end{equation}
These two quantities are shown in Fig.~\ref{fig:Iamp} and Fig.~\ref{fig:Samp} respectively.  Ignoring the fine structure, the general qualitative behavior for the current shows a broad maximum around $V=\epsilon$ (corresponding here to $V=0.6\Delta$), and a slow decrease at lower voltage. This scale corresponds to the voltage where the energy level of both dots coincides with the chemical potential of the lateral electrode each of them is connected to. Note that the current amplitude is at most of the order of 1 (in units of $\Delta$) and decreases to $\sim 0.1$ for the smallest voltages considered. The noise amplitude, however, shows a radically different behavior, with sudden bursts which become more marked as the voltage gets smaller. Remarkably, while these fluctuations can make the noise reach values as high as $\sim 100$ (in units of $\Delta$), there are some wide regions of voltage (for example near $V=0.3 \Delta$) where the noise amplitude is several orders of magnitude smaller.  Note that the details of the amplitude of the noise vary importantly when the parameters $\epsilon$ and $\Gamma$ are varied (even within the resonant dots regime). Huge noise varying strongly with voltage has been observed~\cite{prl19_jonckheere} by some of the authors in a different context (a system of three one dimensional topological superconductors which bear Majorana fermions at their extremities) and has been attributed to a different underlying physical mechanism: the presence of a zero Majorana mode.

\section{Discussion}
\label{sec_discussion}

One of the main goals of this application of the present formalism has been to address whether, upon decreasing the bias voltage, the noise signal of MCPR decreases, and goes to zero despite a non equilibrium situation, in the same manner that a DC Josephson junction can bear only thermal fluctuations at equilibrium. In this sense, our results seem to be in qualitative agreement with Ref.~\onlinecite{prb93_melin}, which deals with a three superconducting lead device, while also surpassing it for the following reasons.

First, the formalism we have developed here is general enough to allow to treat the non equilibrium current and noise characteristics of a large class of mesoscopic systems (as exposed in the first part of this paper). Ref.~\onlinecite{prb93_melin} focused on a given geometry - a central dot system connected to all three superconducting leads - and all presented calculations were specific to this context. Moreover, the geometry chosen in Ref.~\onlinecite{prb93_melin} does not optimize CAR processes, which are indispensable for the optimal detection of MCPR. In past works on the CPBS,\cite{prb83_chevallier,prb85_rech} the authors showed that the optimization of CAR processes is achieved by considering a device where the two quantum dots connected to the normal metal leads have energy levels which are opposite with respect to the central grounded superconductor. This motivated, concerning the study of the ASCPBS\cite{prb87_jonckheere} the same choice of an antisymmetric dot configuration, so that direct Andreev processes are also filtered out. In this respect, the present work achieves a more rigorous diagnosis of noise for MCPR because CAR processes are optimized from the start.   

Second, our approach is non-perturbative, taking into account the coupling to/from dots and leads to all orders, for an arbitrary superconducting device. The authors of Ref. ~\onlinecite{prb93_melin}, however, resorted to a truncated perturbative expansion to compute the transport properties. Beyond the obvious inherent limitations of perturbation theory when it comes to dealing with high junction transmission or resonant situations, such an approach also imposes the introduction of a phenomenological dot parameter intended to regularize the perturbative treatment. This small parameter corresponds to a finite intrinsic linewidth broadening which needs to be systematically added ``by hand''  to the bare dot Green's function from the start. The bare dot constitutes an isolated, coherent system, and adding such an infinitesimal in the bare dot Green's function cannot be justified from a rigorous microscopic model. It is typically introduced to simulate the coupling to a (dissipative) electrodynamic environment, to describe a source of decoherence for the quantum dots. The addition of such an extra phenomenological parameter may have unforeseen repercussions on the transport properties at a MCPR, which were not thoroughly explored in Ref.~\onlinecite{prb93_melin}, casting a shadow on the generality of their results. In our present work, an infinitesimal parameter enters the advanced/retarded lead Green's function only, and this infinitesimal is of course rendered finite for numerical calculations. This is the norm, and subtleties about the interplay between this infinitesimal  and the (adiabatic) voltage have been addressed  in Ref. \onlinecite{prb54_cuevas}. We checked systematically that the current and noise computed within our formalism were not affected when this infinitesimal (associated with the lead Green's functions) was further reduced, converging toward the same results, while the bare dots Green's functions are systematically treated as fully coherent entities, devoid of any phenomenological linewidth for the dot system. 

\section{Conclusion}\label{sec_conclusion}

We have studied a general class of multi-terminal superconducting devices composed of superconducting and normal metal leads arbitrarily connected to a quantum dot system. Expressions for the non equilibrium current and noise have been cast into matrix products of self energies and dressed single particle Green's function. The numerical solution of the latter via Dyson's equation allows to characterized the transport properties of a vast class of systems. This work constitutes a Hamiltonian formulation of quantum transport in the same spirit as the Landauer-Buttiker-Imry multichannel/multi-terminal formalism,\cite{prb31_buttiker,prl57_buttiker,prb45_martin,prb46_buttiker} extended to superconducting hybrid systems. 

We applied our formalism to the ASCPBS, a three-terminal superconducting device designed with a central grounded electrode contacted to two other leads via two QD nanowires. When CAR processes are operating on the central superconductor and the voltages applied to the lateral leads are commensurate, the partial currents are known to depend on the bare superconducting phase differences, leading to a Josephson-like signature, albeit in out-of-equilibrium conditions.~\cite{prl106_freyn,prb87_jonckheere} This has been confirmed by experimental signatures for the differential conductance~\cite{prb90_pfeffer} and more recently explored for the noise.~\cite{pnas_cohen} Such a phenomenon is referred to as a MCPR since the underlying explanation involves the correlated motion of several CPs between the three superconducting reservoirs, with split Cooper pairs from the central lead. 

Within the out-of-equilibrium Keldysh framework, the equivalent of a path integral approach leads to a Dyson equation which relates bare and dressed QD Green's functions, introducing a self-energy term which accounts for the coupling to superconductors in a non-perturbative way. The statistics of the current operator have been derived using counting fields~\cite{prb83_chevallier} which allow generalizations to full counting statistics in principle. The commensurability of the voltages, which is assumed here for computational reasons and because it is a requirement for the observation of MCPR, allows the use of a double Fourier transformation leading to a convenient matrix representation. As a result, all observables of this system (current, noise autocorrelations and noise crossed correlations), are expressed in terms of the dressed Green's function of the dots, a procedure which requires a (large) matrix inversion of the Dyson equation. 

More specifically, we have focused on the quartet resonance where the voltages imposed on the two lateral leads are opposite, and two regimes have been numerically investigated. In the QPC junction limit, where the dot energy levels are placed outside the gap of the central superconductor, and energy filtering of the dots is not effective, we have found that the noise correlations decrease when the voltage is lowered, so that the Fano factors take very low values in the adiabatic limit. We have also demonstrated that the decrease of the Fano factor is not specific to the quartet resonance, by showing similar behaviors for three other voltage configurations.  

This observation is in sharp contrast with the case of a two-terminal junction where the Fano factor varies as $\sim 1/V$ at low voltage. In this regime, negative noise crossed correlations are generally observed at all voltages, except for fine-tuned values of the quartet phase.
In the resonant dots regime, where the dot energy filtering is sharper and occurs within the gap of the central superconductor, only positive values of the noise crossed correlations are obtained, an indication of the dominance of the quartet process. Compared to the QPC regime, large values of the noise have been obtained together with a strong sensitivity on both phase and voltage due to the acute energy filtering. The appearance of giant Fano factors constitutes an important signature of this regime.

We therefore believe that the present noise diagnosis of our 3-superconductor setup sheds further light on the physics of multiple Cooper pair processes, with potential repercussions on experimental detection. 

Possible extensions of this work could follow several directions including 

(i) the investigation of lower voltages in the resonant dots regime which would require more efficient integration tools in order to further probe the noise reduction; 

(ii) a thorough analysis of other voltages associated with MCPR ($mV_a+nV_b=0$), for larger integers $m$ and $n$, which differ from the specific quartet process studied in this application of our formalism. Higher fractions have been extensively studied, albeit only for the DC Josephson current, in both non-equilibrium and equilibrium setups:\cite{prb87_jonckheere,prb90_rech} to each higher order MCPR corresponds an Andreev bound state described by a combination of CAR and direct Andreev processes in the ASCPBS which describe a closed orbit in energy space. Basic transport processes are the same, but they involve more CAR processes necessarily. We have shown that, in the QPC regime, the Fano factor decreases for voltage configurations other than the quartet resonance. While the behavior is qualitatively the same for all voltage configurations, a detailed exploration of the latter could provide precious insight on these systems. 

(iii) the influence of inter-dot tunneling, which may spoil the efficiency of CAR processes in the ASCPBS: this was indeed the case for the CPBS in Ref. \onlinecite{prb83_chevallier}, and we expect the MCPR signal to be decreased in the presence of such tunneling. On general grounds, strong inter-dot tunneling should be taken into account in a general, large quantum dot ``array'', with the goal to describe a specific mesoscopic device or molecule with many sites/orbitals. 

(iv) the inclusion of different injection regions in the leads when a superconductor is connected to several quantum dots. We aware that dimensionnality and $k_F$ oscillations can lead to a modification of the coupling parameters that we have used here, in the same spirit as in Ref. \onlinecite{prb92_jacquet}, where the reduction of CAR over large points of injections was taken into account effectively.

(v) the influence of the QD Coulomb on-site energy, which could be done using either a Hubbard-Stratonovich treatment (followed by a standard approximation such as a saddle point method~\cite{prb92_jacquet}) or a self-consistent perturbative approach;~\cite{prb85_rech}

(vi) the investigation of finite frequency noise at multiples of the Josephson frequency (as well as the Josephson frequency current harmonics), for which we have available analytic expressions, but which would require a challenging experimental detection scheme similar to that of Shapiro steps. Particularly relevant would be to study the effect of decoherence (due to the coupling of the dot system to a normal metal lead for instance)  on the amplitude of the higher noise harmonics. 

(vii) the investigation of electron-hole decoherence effects on the zero frequency noise, using methods borrowed from quasiclassical circuit theory~\cite{nazarov,nazarov_blanter_2009,prb95_padurariu}.

\acknowledgments
We acknowledge the support of the French National
Research Agency, through the project ANR NanoQuartets
(ANR-12-BS1000701). Part of this work
has been carried out in the framework of the Labex
Archimède ANR-11-LABX-0033. The project leading to this publication has received funding from Excellence Initiative of Aix-Marseille University - A*MIDEX, a French ``Investissements d'Avenir'' programme.

\bibliography{biblio}

\end{document}